\documentclass[10pt, a4paper]{amsart}

\usepackage[all]{xypic,xcolor}

\usepackage{url,amssymb,slashed}

\newcommand{\Hom}[2]{\mathrm{Hom}\left({#1},{#2}\right)}
\newcommand{\Z}{\mathbb{Z}}
\newcommand{\R}{\mathbb{R}}
\newcommand{\C}{\mathbb{C}}

\newcommand{\pt}{\text{pt}}

\newcommand{\ud}{\mathrm{d}}

\newcommand{\diffEl}[1]{\check{#1}}
\newcommand{\Forms}{\Omega}
\DeclareMathOperator{\Curv}{Curv}
\DeclareMathOperator{\ch}{Ch}

\newcommand{\cmap}[1]{c\left(#1\right)}
\newcommand{\formsmap}{{i}}
\newcommand{\diffGroup}[1]{\check{#1}}

\newcommand{\even}{\mathrm{ev}}

\newcommand{\mukai}[2]{\left\langle{#1},\,{#2}\right\rangle}
\newcommand{\Lie}{\mathcal{L}}
\newcommand{\orth}{\mathfrak{o}}
\newcommand{\gl}{\mathfrak{gl}}
\newcommand{\spin}{\mathfrak{spin}}
\DeclareMathOperator{\Spin}{Spin}
\DeclareMathOperator{\Cliff}{Cliff}
\newcommand{\Dirac}{\slashed{D}}
\newcommand{\Ahat}{\hat{A}}

\newcommand{\tr}{\mbox{Tr\,}}

\newcommand{\calL}{\mathcal{L}}

\newcommand{\beq}{\begin{equation}}
\newcommand{\eeq}{\end{equation}}
\newcommand{\bea}{\begin{eqnarray}}
\newcommand{\eea}{\end{eqnarray}}

\newcommand{\dd}{\mbox{d}}

\makeatletter
\@addtoreset{equation}{section}
\makeatother

\newtheorem{theorem}{Theorem}[section]

\newtheorem{proposition}[theorem]{Proposition}
\theoremstyle{definition}

\theoremstyle{remark}

\theoremstyle{remark}
\newtheorem{example}[theorem]{Example}
\numberwithin{equation}{section}

\begin{document}
\rightline{\small IPhT-t13/018}
\vskip 1cm

\title{D-brane couplings and Generalised Geometry}
\author{Alexander Kahle}
\address{Max Planck Institute for Mathematics, Bonn}
\email{kahle@uni-math.gwdg.de}
\author{Ruben Minasian}
\address{Institut de Physique Th\'eorique, CEA/Saclay}
\email{ruben.minasian@cea.fr}
\begin{abstract}
The goal of this paper is to re-examine $D$-brane Ramond-Ramond field couplings in the presence of a $B$-field. We will argue that the generalised geometry induced on the world volume by the $B$-field results in an important but subtle change on the coupling. In order to explain this, we use the language of differential $K$-theory. The expression determining the coupling is then seen to be a consequence of the Riemann-Roch theorem. Our key assertion is that the appropriate Riemann-Roch theorem changes in the presence of the $B$-field.  In particular, the A-hat forms appearing in the theorem are now constructed using the \emph{torsionful} Levi-Civita connection associated to the generalised geometry. As we shall see, the resulting expression not only agrees with recently discovered local couplings on the $D$-brane worldvolume involving RR fields and  derivatives of the $B$-field, but also makes the coupling manifestly $T$-duality invariant.
\end{abstract}
\maketitle
\section{Introduction}
The goal of this paper is to re-examine $D$-brane Ramond-Ramond field couplings in the presence of a $B$-field. We will argue that the generalised geometry induced on the world volume by the $B$-field results in an important but subtle change in the coupling, which, when properly understood, makes it $T$-duality invariant.

The basic framework for discussing $D$-branes and Ramond-Ramond fields has long been understood \cite{Fr:Dirac}: Ramond-Ramond fields are generalised self-dual abelian gauge fields where the charge carrying objects, the $D$-branes, carry charges quantised by certain \emph{$K$-theory} groups. 

The involvement of $K$-theory in a proper account of Ramond-Ramond fields was first realised in  \cite{MiMo} by examining the $D$-brane couplings, and notably the appearance of the  square root of the A-hat genus in the coupling between the Ramond-Ramond fields and $D$-brane.  The realisation that $D$-branes are in fact charged by $K$-theory ($K^0$ for type IIB, $K^1$ for type IIA) eventually lead to  the proper framework for understanding the theory is \emph{differential $K$-theory}. Setting the theory in this framework automatically produces the initially puzzling ``A-hat" factor as a consequence of Riemann-Roch theorem. 

The presence of $B$-fields complicates the story:    the $B$-field \emph{twists} the $K$-theory group quantising $D$-brane charges \cite{Witten1998, BM}. Ramond-Ramond fields are now \emph{$B$-twisted} differential $K$-theory co-cycles and the coupling with $D$-branes needs to take this into account. Impressionistically, the coupling is of the following form:
\begin{equation}\label{eq:dbraneimpression}
(constant)\int_W\sqrt{\frac{\Ahat(TW)}{\Ahat( \nu)}}e^{-B}i^*C\wedge\ch V,
\end{equation}
where $i:W\to X$ is the support of the $D$-brane $W$ in the bulk $X$, $\Ahat(TW)$ and $\Ahat(\nu)$ are, respectively, A-hat forms for the tangent and normal bundles to $W$. $C$ is a local polyform representing the Ramond-Ramond field potential, $B$ is a local two-form representing the $B$-field potential, and $V$ is the Chan-Paton bundle. We have suppressed an additional factor that would appear if the $D$-brane $W$ were only $\text{spin}^c$ and not spin. 

The general form \eqref{eq:dbraneimpression} has long been known, but the precise coupling remains elusive. In our paper we focus in on one particular source of ambiguity -- the explicit form of the A-hat terms -- and hope to resolve it. In doing so, we also explain terms in the Ramond-Ramond field/$D$-brane coupling predicted in recent work that seem puzzling in a naive interpretation of Eq.~\eqref{eq:dbraneimpression}~\cite{BeckRob, Ga, GaM}. In this work, perturbative methods are used to argue that terms involving second derivatives of the $B$-field should appear. Perhaps even more puzzling, it is suggested that there should be terms involving contractions of the Ramond-Ramond field with derivatives of the metric and of the $B$-field!

Our thesis is as follows: \emph{the $B$-field induces a generalised geometry on the bulk and on the $D$-brane}.\footnote{In a slightly different context, this point of view was explored in \cite{CoStriWa} to explain certain aspects of the action in supergravity.} This has several effects, the salient one for us being that \emph{Levi-Civita connection gets replaced by a torsionful connection} with the torsion proportional to the $B$-field fieldstrength. As already noted by Bismut in \cite{BisTor}, the local Riemann-Roch theorem changes when the Levi-Civita connection is replaced by a connection with totally skew torsion: the A-hat form now gets built out of the curvature of a connection with the \emph{opposite} torsion! We argue that this in turn changes the form of pairing between twisted differential $K$-theory classes, accounting for the factors involving the derivative of the $B$-field fieldstrength observed by \cite{BeckRob, Ga, GaM}.

Before going on to describe this in more detail, perhaps we should pause to examine why the precise form of the A-hat terms matters. After all, general arguments show that changing the connection used to build the A-hat form changes it by an exact form, and one may thus naively expect the integral \eqref{eq:dbraneimpression} to be unaffected by the such changes. However, the usual Stokes theorem argument fails! The Ramond-Ramond potential is not closed, and thus the integral is sensitive to any change of the integrand, even exact changes.

Returning to our situation: we argue that the presence of the $B$-field changes the Levi-Civita connection $\nabla^{LC}$ to a new connection $\nabla^H=\nabla^{LC}+\frac{1}{2} H$,\footnote{We describe below how this connection arises naturally from the generalised geometry induced by the $B$-field.} where $H=\ud B$ is the $B$-field fieldstrength. Bismut then tells us that in the local index theorem, $\Ahat(\Omega^{LC})$ gets replaced by $\Ahat(\Omega^{-H})$. As a result, we are led to a pairing on the bulk of the form 
\[
(constant)\int_X\mukai{C}{\sqrt{\Ahat(\Omega^{-H})}e^{-B}i_*\ch V},
\]
where $\mukai{A}{B}$ is the Mukai pairing.
However, this pairing is not symmetric in its arguments!\footnote{The pairing is required to be symmetric for various reasons, not least in order to be amenable to quantisation.} Upon symmetrising, this yields a coupling of the form 
\[
\frac{1}{2}(constant)\int_X\mukai{C}{\sqrt{\Ahat(\Omega^{-H})}e^{-B}i_*\ch V}+\mukai{\sqrt{\Ahat(\Omega{^H})}e^{B}i_*\ch V}{\bar{C}}.
\]
This coupling has several features, the most notable being that it is ``even" in $H$, agreeing with the symmetry of the theory under the reversal of the $B$-field as well as the need to have only CP-even couplings\footnote{The bulk counterparts and the higher derivative corrections in type II theories  will be discussed in \cite{LiuM}.}. While the Ramond-Ramond fields and the $D$-brane currents  have a $B$-twist, the Mukai pairing makes sure that the coupled fields are ``zero-$B$"! This is important, as $B$-twisted differential forms may not sensibly be integrated -- changing a $B$-twisted differential form by a $B$-exact differential form changes the integral. Physically, this means that the coupling has the \emph{correct} gauge invariance. This ``zero-$B$" structure, and its importance, becomes manifest when interpreting Ramond-Ramond fields as cocycles in $B$-twisted differential $K$-theory, and the coupling as a pairing taking place in that setting. We develop the theory in Sec.~\ref{sec:twist}, and we argue that the coupling above should in fact be interpreted as the Chern character of the ``true" coupling in differential $K$-theory.

In order to make contact with the predictions of \cite{BeckRob, Ga, GaM}, we need to pull back the coupling to $W$.  A naive comparison shows that our coupling, when linearised, reproduces the ``non-contracted" terms, but seems not to predict the remaining terms. Where do these come from? In Sec. \ref{sec:tdual}, we argue that these are in fact already there, and are due to the failure of $\Ahat(\Omega^{\pm H})$ to split into tangent and normal parts when pulled to the $D$-brane worldvolume. Finally the parity of the $B$-field (the contracted terms with even/odd number of contracted indices contain only even/odd powers of $H$) are made manifest when one examines the $T$-duality properties of the couplings.

The remainder of the paper is organised as follows: we review the necessary mathematical background in Sec.~\ref{sec:mathbackground}. Secs.~ \ref{sec:gengeom} and \ref{sec:diffcoh} are reviews of well-known material, respectively  generalised geometry and differential $K$-theory. Sec.~\ref{sec:twist} is new, and in some sense provides the mathematical foundation  of the paper. Here we propose that the torsionful connection arising from the generalised geometry induced by the $B$-field be used to construct the bilinear pairing on $B$-\emph{twisted} differential $K$-theory, and relate its Chern character to the Mukai pairing. Sec.~\ref{sec:maxwell} discusses generalised Abelian gauge theories from a mathematical point of view, and frames our discussion of the pairing between Ramond-Ramond fields and $D$-branes in the presence of a $B$-field. Secs.~\ref{sec:rrfield} and ~\ref{sec:tdual} form the heart of our paper. In the first, we derive our proposal for the Ramond-Ramond/$D$-brane coupling. We see that the twisted ``A-hat" form arises naturally from Riemann-Roch in this context. Sec.~\ref{sec:tdual} is more ``physicsy": here we give an argument based on $T$-duality explaining the ``contracted" terms predicted by \cite{BeckRob, Ga, GaM}. We see that the parity of the $B$-field in the coupling (even or odd) is automatically predicted by our argument. Finally, we discuss further avenues of research. 

\subsection*{Acknowledgements} We would like to thank Simons Center for Geometry and Physics (AK and RM), the Isaac Newton Institute (RM) and the Max Planck Institute for Mathematics (AK) for hospitality during the course of this work. We thank K. Becker, G. Guo, and especially D. Robbins for patient explanations of their work;  useful discussions with J. Liu  and C. Scrucca are also gratefully acknowledged. The work of RM is supported in part by ANR grant 12-BS05-003-01. The work of AK was partially supported by a grant  from the German Research Foundation (Deutsche Forschungsgemeinschaft (DFG)) through the 
Institutional Strategy of the University of G\"ottingen.

\section{Mathematical background}\label{sec:mathbackground}
\subsection{Hitchin's generalised geometry}\label{sec:gengeom}
Generalised geometry, first introduced by Hitchin and Gualtieri (and nicely reviewed in e.g.~\cite{Gualt, Hi}), springs from a basic observation: for any smooth manifold $M$ of dimension $n$, there is a natural indefinite inner product of signature $(n,n)$ on the bundle $T(M)\oplus T^*(M)$ (henceforth abbreviated $T\oplus T^*$) defined by
\[
(X+\eta,\,X'+\eta')=\frac{1}{2}(i_X\eta'+i_{X'}\eta),
\]
where $X,X'\in T(M)$, $\eta,\eta'\in T^*(M)$. This indefinite product reduces the structure group of $T(M)\oplus T^*(M)$ from $GL(2n)$ to $SO(n,n)$.

The subalgebra bundle $\orth(T\oplus T^*)$ of $\gl(T\oplus T^*)$ is given pointwise by matrices of the form 
\[
\begin{pmatrix}
A&\beta\\
B&-A^t
\end{pmatrix}
\] 
where $A\in \gl(T\oplus T^*)$, $B\in\bigwedge^2T^*$ and $\beta\in\bigwedge^2T$. Thus in addition to transformations coming from the general linear bundle of $T$, there are transformations generated by two-forms and bi-vectors. We call the former $B$-field transforms, and the latter $\beta$-field transforms. Note that, for a $B\in\bigwedge^2T^*$,
\[
\begin{pmatrix}0&0\\B&0\end{pmatrix}^2=0,
\]
so that 
\[
\exp(B)(X+\eta)=X+\eta+i_XB.
\]
There is a similar formula for the action of bi-vectors.

Generalised geometry has one more basic structure: the \emph{Courant bracket}. it essentially plays the role of the Lie bracket on the tangent bundle in ordinary geometry. Given two sections of the generalised tangent bundle, $X+\eta,X'+\eta'\in\Gamma(T\oplus T^*)$, the Courant bracket is defined by the formula
\[
[X+\eta,X'+\eta']=[X,Y]+\Lie_X\eta'-\Lie_{X'}\eta-\ud(i_X\eta'-i_{X'}\eta).
\]
The Courant bracket breaks the symmetry between $B$ and $\beta$-field transforms: the Courant bracket commutes with $B$-field transformations generated by \emph{closed} two-forms, while it commutes with \emph{no} $\beta$-field transformations. Indeed, for a closed two-form $B$, one calculates 
\[
[\exp(B)(X+\eta),\exp(B)(X'+\eta')]=[X+\eta,X'+\eta']+i_{[X,X']}B=\exp(B)[X+\eta,X'+\eta'].
\]
Clearly, diffeomorphisms also commute with the Courant bracket via pullback.

This brings us to the key point: in generalised geometry, i.e.~the geometry of $T\oplus T^*$ along with the pairing $(\cdot,\cdot)$ and the Courant bracket, the group of symmetries is extended from the diffeomorphism group of $X$ by the closed two forms. \emph{The basic group of symmetries of generalised geometry is $\Forms_{\ud=0}^2(M)\rtimes \mathrm{Diff}(M)$.}

The Lie algebra of $\Forms_{\ud=0}^2(X)\rtimes \mathrm{Diff}(M)$ is generated by closed two forms $B\in\Forms^2_{\ud=0}(M)$ and vector fields $X\in\Gamma(T)$. Taking $B=\ud\eta$, $\eta\in\Omega^1(M)$, we see that the Lie Algebra action of $X+\ud\eta$ on $X'+\eta'$ is given by
\[
(X+\ud\eta)\cdot(X'+\eta')=\Lie_X(X'+\eta')-i_X'\ud\eta=[X,X']+L_X\eta'-L_{X'}\eta+\ud(i_{X'}\eta).
\]
Skew symmetrising the right hand side gives a new interpretation of the Courant bracket: 
\[
[X+\eta,X'+\eta']=\frac{1}{2}\left[(X+\ud\eta)\cdot(X'+\eta')-(X'+\ud\eta')\cdot(X+\eta)\right].
\]
Note, however, that the Courant bracket is not the Lie bracket in the Lie algebra of $\Forms_{\ud=0}^2(X)\rtimes \mathrm{Diff}(M)$ (it involves \emph{one-forms}, not two-forms) and does \emph{not} satisfy the Jacobi identity. Instead, it satisfies a sort of derived Jacobi identity: 
\[
[[u,v],w]+[[v,w],u]+[[w,u],v]=\frac{1}{3}\ud\left(([u,v],w)+([v,w],u)+([w,u],v)\right)
\]
for $u,v,w\in\Gamma(T\oplus T^*)$. The Courant bracket has two other characteristic properties:
\begin{align*}
[u,fv]&=f[u,v]+(Xf)v-(u,v)\ud f,\\
X(v,w)&=([u,v]+\ud(u,v),w)+(v,[u,w]+\ud(u,w)),
\end{align*}
where $u+X+\eta,v,w\in\Gamma(T\oplus T^*)$. 
\subsubsection{Spinors and the Mukai pairing}
Just as in ordinary geometry, it is profitable to ask when one might lift the $SO(n,n)$ structure to a $\Spin(n,n)$ structure on $T\oplus T^*$. It turns out the sole obstruction is orientability of $M$, and upon choosing and orientation for $M$, the $SO(n,n)$ structure lifts. The differential forms play a special role in generalised spin geometry, as we shall now see.

There is a natural action of sections of $T\oplus T^*$ on the differential forms: for $X+\xi\in T\oplus T^*$ and $\omega\in\Forms^*(X)$, one defines
\[
(X+\xi)\cdot\omega=i_X\omega+\xi\wedge\omega.
\]
This action lifts to an action of the Clifford algebra bundle of $T\oplus T^*$ with its $O(n,n)$ structure, $\Cliff(T\oplus T^*)$. Indeed, one verifies that
\[
(X+\xi)^2\cdot\omega=i_X(\xi\wedge\omega)+\xi\wedge(i_X\omega)=(i_X\xi)\wedge\omega=(X+\xi,X+\xi)\omega.
\]
Thus the differential forms on $M$ naturally form a $\Cliff(T\oplus T^*)$ module. 

One might suspect that the exterior bundle $\bigwedge^*T$ in fact forms the bundle of spinors for $T\oplus T^*$, but this is not quite the case. In fact, the bundle of spinors is the bundle
\[
S=\bigwedge^*T^*\otimes(\bigwedge^nT^*)^{-1/2}
\]
and a choice of nowhere vanishing half-density on $M$ is needed to identify the spinors with the differential forms. There is always a bilinear pairing on $O(n,n)$ spinors, which, from the above decomposition, gives a canonical bilinear pairing from forms into the determinant bundle of $T$ called the \emph{Mukai pairing}. Explicitly, given two differential forms $\omega_1$, $\omega_2\in\Forms(M)$, the pairing is given by
\[
\langle \omega_1,\omega_2\rangle=\sum_i(-1)^i\left(\omega_1^{2i}\wedge\omega_2^{n-2i}+\omega_1^{2i+1}\wedge\omega_2^{n-2i-1}\right),
\]
where $\omega^i$ denotes the part of the differential form in degree $i$. 

There is a natural action of two forms on spinors (and differential forms): differential forms embed into $\Cliff(T\oplus T^*)$ in the standard way (this is not an algebra map!) and a given two-form $B\in\Forms^2(M)$ via this embedding -- 
\[
B\cdot\omega=B\wedge\omega.
\]
In fact, one may see that the two forms acting this way lie inside $\spin(T\oplus T^*)\subset\Cliff(T\oplus T^*)$, and thus exponentiate to a $B$-field action on spinors:
\[
\exp B\cdot\omega=e^{-B\wedge}\omega
\]
(note the sign!). This, along with the usual diffeomorphism action gives an action of $\Forms^2_{\ud=0}\rtimes\mathrm{Diff}(M)$ on spinors and differential forms (which, because it acts, preserves the Mukai pairing). In particular, recalling that sections of $T\oplus T^*$ embed into the Lie algebra of $\Forms^2_{\ud=0}\rtimes\mathrm{Diff}(M)$ via $X+\eta\mapsto X-\ud\eta$ one obtains a ``Lie derivative" on differential forms $\omega\in\Forms(X)$
\[
\Lie_u\omega=\ud(u\cdot\omega)+u\cdot\ud\omega,
\]
for $u\in\Gamma(T\oplus T^*)$.
\subsubsection{(Pseudo)-Riemannian Geometry}
As would be expected, introducing a metric on $T$ has implications on the generalised geometry of $M$. Let $g:S^2(T)\to\R$ be a metric on $T$ of signature $(p,n-p)$. This corresponds to reducing the structure group of $T$ to $O(p,n-p)$ and we shall see that it introduces a corresponding reduction of the structure group of the generalised tangent bundle $T\oplus T^*$ to $S(O(p,n-p)\times O(n-p,p))$. A good way to see this is by regarding the metric as a map from $T$ to $T^*$ by
$X\mapsto g(X,\cdot)$. We then have two canonical sub-bundles of $T\oplus T^*$: the graph of $g$, which we denote $V_g$, spanned elements of the form $X+g(X,\cdot)$; and the graph of $-g$, denoted $V_{-g}$, spanned by elements of the form $X-g(X,\cdot)$. It is easy to see that the $O(n,n)$ inner product restricted to $V_g$ (resp.~$V_{-g}$) reduces to $g$ (resp.~$-g$) so that $V_g$ and $V_{-g}$ are bundles with structure groups $O(p,n-p)$ and $O(n-p,p)$ respectively. Indeed, one computes on $V_{\pm g}$
\begin{align*}
(X\pm g(X,\cdot),Y\pm g(Y,\cdot))&=\frac{1}{2}(i_X(\pm g(Y,\cdot))+ i_Y(\pm g(X,\cdot)))\\
&=\pm\frac{1}{2}(g(Y,X)+ g(X,Y))\\
&=\pm g(X,Y).
\end{align*}
Both $V_{g}$ and $V_{-g}$ are canonically isomorphic to $T$, and we denote the lift of a vector field $X\in\Gamma(T)$ to $V_{\pm g}$ by $X_\pm$. These may be written in terms of the orthogonal projections $\pi_{V_{\pm g}}:T\oplus T^*\to V_{\pm g}$. Indeed,
\[
\pi_{V_{\pm g}}X=\frac{1}{2}\pi_{V_{\pm g}}(X+g(X,\cdot)+X-g(X,\cdot))=\frac{1}{2}X_\pm.
\]
There is a very natural expression for the Levi-Civita connection in terms of the Courant bracket and the lifts to $V_{\pm g}$ given by the following proposition (proved in e.g.~\cite{Hi} Prop.~3):
\begin{proposition}\label{prop:levicivita}
Let $v\in\Gamma(V_g)$ and $X\in\Gamma(T)$. Then
\[
\nabla_Xv=\pi_V[X_-,v]
\]
defines a torsionfree connection on $v$ preserving the inner product induced from $T\oplus T^*$ which we call the Levi-Civita connection. 
\end{proposition}
The name is justified as computation shows that the connection $\nabla$ is the ``lift" of the Levi-Civita connection on $T$ to $V_g$. Of course there is also an analogous Levi-Civita connection defined on $V_{-g}$. 

The choice of a metric on $T$ also gives a canonical section of $\Lambda^nT^*$, and as a result a canonical isomorphism between differential forms and generalised spinors, i.e.~sections of the bundle $S=\Lambda^*T^*\otimes(\Lambda^nT^*)^{-1/2}$.

\subsubsection{Generalised tangent bundles}
A crucial feature of the machinery discussed so far is that it is  invariant under an extension of the diffeomorphism group by closed two forms. This invariance allows us to generalise away from the geometry of $T\oplus T^*$, and it is this generalisation that is the real interest for us.

Suppose now that $M$ has a good cover $\left\{U_\alpha\right\}$, and we have a collection of closed two-forms $B_{\alpha\beta}\in\Forms^2_{\ud=0}(U_\alpha\cap U_\beta)$ satisfying the co-cycle condition
\[
B_{\alpha\beta}+B_{\beta\gamma}+B_{\gamma\alpha}=0.
\]
Using the $B$-field action on the double intersections, we may then define an extension of $T$ by $T^*$, $E$ -- a rank $2n$ vector bundle with Courant bracket and $O(n,n)$ inner product
\begin{equation}\label{eq:gentanexact}
\xymatrix{
0\ar[r]&T^*\ar[r]&E\ar[r]^\pi&T\ar[r]&0.
}
\end{equation}
Such an extension is called an exact Courant algebroid, and every exact Courant algebroid may be constructed from such a collection of closed two-forms.

Concretely, on each $U_\alpha$ we identify $E|_{U_\alpha}$ with $T\oplus T^*$ and use the Courant bracket and inner product discussed above. We then use $B$-field transformations to patch: two sections $u_\alpha\in\Gamma(T\oplus T^*|_{U_\alpha})$, $u_\beta\in\Gamma(T\oplus T^*|_{U_\beta})$ patch to give a section of $u\in\Gamma(E|_{U_\alpha\cup U_\beta})$ when
\[
u_\beta=e^B\cdot u_\alpha.
\]
In a similar way, the $B$-field action can be used to twist the complex of differential forms to a $\Z/2\Z$-graded complex acted on by $\Cliff(E)$, which we will call the $B$-twisted differential forms.

Exact Courant algebroids determine a class in $H^3(M,\R)$, which we will denote by $[E]$. The set of closed two-forms $B_{\alpha\beta}$ is a one-cocycle for the sheaf $\underline{\Forms}_{\ud=0}^2(M)$ which fits into the exact sequence of sheaves
\[
\xymatrix{
0\ar[r]&\underline{\Forms}_{\ud=0}^2(M)\ar[r]&\underline\Forms^2(M) \ar[r]^\ud&\underline{\Forms}_{\ud=0}^3(M)\ar[r]&0.
}
\]
All the sheaves in the sequence are flabby, so that
\begin{gather*}
H^1(M,\underline{\Forms}^2_{\ud=0}(M))\\\simeq H^0(M,\underline{\Forms}^3_{\ud=0}(M)))/\ud H^0(M,\underline{\Forms}^2(M))=   \Forms^3_{\ud=0}(M)/\ud\Forms^2(M)=H^3(M,\R).
\end{gather*}
In fact, refining the above argument shows that there is a unique closed three-form $H$ determined by the $B_{\alpha\beta}$ such that $[H]$ represents the characteristic class of $E$. The complex of $B$-twisted differential forms is equivalent to the complex $(\Forms^\bullet(M),\ud+H)$.

As we shall see in our discussion of differential co-cycles, a differential 3-cocycle $\check{B}\in\diffGroup{\mathcal{H}}^3(M)$ determines a set of $F_{\alpha}\in\Forms^2(M)$ such that $\ud F_\alpha=\Curv\check{B}|_{U_\alpha}$. Thus one may define $B_{\alpha\beta}=F_\beta-F_\alpha$ and easily check that $\ud B_{\alpha\beta}=0$ and that the $B_{\alpha\beta}$ form a one-cocycle. A differential three-cocycle thus \emph{defines} a generalised tangent bundle $E_{\check{B}}$.  In this case the characteristic class associated to $E_{\check{B}}$ will be in the image of integral cohomology in real cohomology. In fact $[E_{\check{B}}]=c(\check{B})$ in $H^3(M;\R)$. The complex twisted differential forms in this case are equivalent to the $\Curv{\check{B}}$-twisted differential form. In fact, one obtains more: the two-forms $F_\alpha$ give a canonical isotropic splitting of the sequence
\[
\xymatrix{
0\ar[r]&T^*\ar[r]&E_{\check{B}}\ar[r]^\pi&T\ar[r]&0;
}
\]
one sends $X\in\Gamma(T|_{U_\alpha})$ to $X+i_XF_\alpha\in\Gamma(E|_{U_\alpha})$.

Twisted pseudo-Riemannian structures are also easy to describe in this context: they are simply reductions of the $SO(n,n)$ bundle to $S(O(p,n-p)\times O(n-p,p))$. More concretely, they consist of two orthogonal rank-$n$ sub-vector bundles $V_\pm$ of $E$ on which the metric restricts, respectively, to a non-degenerate form of signature $(p, n-p)$ and $(n-p, p)$. Locally on an open set $U_\alpha$, using the isomorphism $E\simeq T\oplus T^*$, one identifies $V_\pm$ as graphs of some $\pm h_\alpha:T\to T^*$. Comparing $h_\alpha$ and $h_\beta$ on $U_{\alpha\beta}$, one sees that the symmetric parts are equal and thus define well-defined global $(p,n-p)$ metric on $M$. The skew parts change: let the skew part of $h_\alpha$ be $F_\alpha$; then one computes that $F_\alpha(X)=F_\beta(X)+i_XB_{\alpha\beta}$. Said differently, the skew parts furnish the data to give a canonical splitting of the sequence \eqref{eq:gentanexact}.

Conversely, a splitting of \eqref{eq:gentanexact} along with a choice of $(p,n-p)$ metric $g$ on $T$ determines the bundles $V_\pm$. In particular, \emph{a differential 3-cocycle $\check{B}$ along with a choice of $(p,n-p)$-metric on $M$ determines a twisted pseudo-Riemannian structure $V^g_{\pm}\subset E_{\check{B}}$.}

A twisted pseudo-Riemannian structure has a canonical Levi-Civita connection associated to it, just as in Prop.~\ref{prop:levicivita}. The presence of a $B$-field induces a key change: the connection on $V$ has \emph{totally skew torsion}! In particular, when $E$ is determined by $\check{B}$, one has that
\[
g(T^\nabla(X,Y),Z)=\Curv\check{B}(X,Y,Z),
\] 
where 
\[
T^\nabla(X,Y)=\nabla_XY-\nabla_YX-[X,Y]
\]
is the torsion of $\nabla$.\footnote{Our derivation of the torsionful Levi-Civita connection follows the original presentation of Hitchin (e.g.~in \cite{Hi}). We have chosen to do so as it is the most direct path to the connection of interest. We should, however, point out that this connection fits into a more general context investigated in \cite{CoStriWa}. In that work, the authors investigate connections on the \emph{entire} generalised bundle (they also include the dilaton in their discussion, but this does not affect the basic issues of concern to us). They formulate a generalised notion of ``torsion" for such connections, and then investigate to what extent torsionfree generalised connections preserving the generalised Riemannian structure  are unique. It turns out that they are not unique: there is a whole family of generalised Levi-Civita connections! This is true even when $M$ is an ordinary pseudo-Riemannian manifold, with the trivial generalised geometry. Fortunately, these connections all agree when restricted to act on certain sub-bundles of the generalised tangent bundle. In particular, the usual Levi-Civita connection may be recovered when restricting the generalised connection to act on the sub-bundles defining the pseudo-Riemannian structure on $M$ when $M$ has the trivial generalised geometry. Allowing $M$ to have non-trivial generalised geometry, and thus a non-trivial $B$-field, one discovers that the usual Levi-Civita connection then deforms to the torsionful connection used in our work (independent of the choice of ``generalised Levi-Civita connection" compatible with the geometry). Thus, even though there is no unique generalised Levi-Civita connection on the entire generalised tangent bundle, one has a canonical generalised Levi-Civita connection (and more importantly for us, Dirac operator) when restriction to the sub-bundles determining the pseudo-Riemannian structure.}
\subsection{Differential cohomology}\label{sec:diffcoh}
Differential cohomology theories were first introduced to mathematics by Deligne \cite{De} and Cheeger and Simons \cite{CheSim} in studying geometric refinements of characteristic classes. The subject, however, gained significant impetus when it was realised that it provides a language perfectly suited to higher (abelian) gauge theory. As we will see, the essential utility in the language lies in the fact that it combines in a non-trivial way the locality of differential forms with the global sensitivity of cohomology classes.

Let $E$ be a generalised cohomology theory (for us, the examples of interest will be ordinary Eilenberg-McLane cohomology and $K$-theory). For a space $X$, there is a canonical map $E^\bullet(X)\to H(X;V)^\bullet$, where $V=E^\bullet(\pt)\otimes\R$.\footnote{$V$ is a $\Z$-graded vector-space, and by $H(-;V)^\bullet$ we denote the total grading of the bi-graded abelian group $H(-;V)$.} On the other hand, the de Rham theorem provides an isomorphism between the cohomology of the complex $\left(\Forms(X;V)^\bullet,\ud\right)$ and $H(X;V)^\bullet$. The $E$-differential cohomology of $X$, $\diffGroup{E}^\bullet(X)$ is defined by the homotopy pullback square \cite{HoSi}:
\[
\xymatrix{
\diffGroup{E}^\bullet(X)\ar[r]\ar[d]&E^\bullet(X)\ar[d]\\
\Forms_{\ud=0}(X;V)^\bullet\ar[r]
&H(X;V)^\bullet
}
\]
It is important to note that the square is \emph{not} cartesian, but a homotopy square, and as a consequence, there are interesting exact sequences\footnote{Strictly speaking, everything said assumes that the ``differential" extensions are extensions as \emph{rings}.} associated to the maps top and left maps:
\begin{gather}
\label{eq:forgetseq}\xymatrix{
0\ar[r]&\left(\frac{\Forms(X;V)^{\bullet-1}}{\Curv\Forms(X;V)^{\bullet-2}}\right)\ar[r]^<<<<<\formsmap&\diffGroup{E}^{\bullet}(X)\ar[r]^c&
E^{\bullet}(X)\ar[r]&
0
},
\\
\label{eq:curvseq}\xymatrix{
0\ar[r]&E^{\bullet-1}(X,\mathbb{R}/\mathbb{Z})\ar[r]&\diffGroup{E}^{\bullet}(X)\ar[r]^>>>>>{\Curv}&\Forms_{\ud=0}(X;V)^\bullet.
}
\end{gather}
The ``forgetful" map $c:\diffGroup{E}^\bullet\to E^\bullet$ is called the \emph{characteristic} map, and the map $\Curv:\diffGroup{E}^\bullet(X)\to\Forms(X;V)^\bullet$ the ``curvature" map. We will soon see the basic example justifying this terminology.

\begin{example}[Cheeger-Simons cohomology] Let us follow Cheeger and Simons and explicitly construct the group $\diffGroup{H}^\bullet(X)$, where $H$ is Eilenberg-MacLane cohomology. Define
\begin{equation}\label{eq:chsimdef}
\diffGroup{H}^k(X)=\left\{\xi\in\Hom{C_{k-1}(X)}{\R/\Z}|\exists\omega\in\Forms^k(X)\forall z\in Z_k\,\xi(\delta z)=\int_z\omega\right\}
\end{equation}
where $Z_k(X)$ (resp.~$C_k(X)$) are the smooth chains (resp.~cycles) on $X$. One may show that for each $\xi\in\diffGroup{H}^k(X)$ there is a unique (and closed) differential form $\omega$ as in Eq.~\eqref{eq:chsimdef}, and that this differential form has integral periods. The map $\Curv:\diffGroup{H}^\bullet(X)\to\Forms^\bullet(X)$ is the assignment $\xi\mapsto\omega$. The characteristic class is given as follows. Let $\tilde{\xi}\in Z^{k-1}(X;\R)$ be such that $\xi(z_{k-1})=\tilde{\xi(z_{k-1})}\mod\Z$ for $z_{k-1}\in C_{k-1}(X)$ (this exists as $\R$ is divisible). By assumption, for any $z_{k}\in C^k$, $\tilde(\xi)(\delta z_k)\mod\Z=\int_{z_k}\omega$, so that there is a $c_k\in Z^k(X;\Z)$ such that $\int_{z_k}\omega=\tilde{\xi}(\delta z_k)+ c_k$. Then $c(\xi)=[c]$. One may check that the assignment is independent of choices.

Let $(L,\alpha)\to X$ be a principal $U(1)$-bundle with connection. The holonomies furnish an element $[L]\in\diffGroup{H}^2(X)$. However, knowing the holonomies of a principal $U(1)$-bundle with connection determines it up to connection-preserving isomorphism, and thus we see that $\diffGroup{H}^2$ is the group of principal $U(1)$-bundles with connection up to connection-preserving isomorphism (the group operation being tensor product). The map $c:\diffGroup{H}^2(X)\to H^2(X)$ assigns to the line bundle its first Chern class and Stokes's theorem shows that $\Curv:\diffGroup{H}^2(X)\to\Forms^2(X)$ assigns ($i/2\pi$ times) the curvature of the connection. These observations justify the naming of the maps $i$ and $\Curv$. We note that the line bundles in the kernel of map $\diffEl{H}^2(X)\to\Forms^2(X)$ are the flat line bundles, and more generally, we will refer to $E^{\bullet-1}(X,\R/\Z)\hookrightarrow \diffGroup{E}^{\bullet}(X)$ as the \emph{flats}. The line bundles with connection in the kernel of the characteristic class map $c:\diffGroup{H}^2(X)\to H^2(X)$ are topologically trivial, and generally we refer to the classes in the kernel of $\diffEl{E}^\bullet(X)\to E^\bullet(X)$ as \emph{topologically trivial}.

In summary, $\diffGroup{H}^2(X)$ gives a geometric refinement of $H^2(X)$: the former classifies principal $U(1)$-bundles up to isomorphism, the latter principal $U(1)$-bundles \emph{with connection}. This example is in some sense paradigmatic. To mention a couple of other examples in the same vein: whereas $H^1(X)$ classifies continuous maps $X\to S^1$ up to homotopy, $\diffGroup{H}^2(X)$ is the group of smooth maps $X\to S^1$; $H^3(X)$ classifies $U(1)$-bundle-gerbes over $X$, $\diffGroup{H}^3(X)$ classifies $U(1)$-bundle-gerbes with \emph{connection and connective structure}.

As a side note: in the Cheeger-Simons model of $\diffGroup{H}^\bullet(X)$, the group structure is clear, but the ring structure is far from easy to define.
\end{example}
\begin{example}[Differential $K$-theory]
As noted earlier, the principal examples that will interest us are when $E=H$, $K$. We now sketch a model of $\diffGroup{K}^0(X)$ due to Simons and Sullivan \cite{SiSu}.\footnote{As is usual in $K$-theory, models for $\diffGroup{K}^i$, $i\ne0$ are less intuitive. The model with possibly the most intuitive cycles is due to Bunke and Schick \cite{BuSchi}. Of course, the machinery of Hopkins and Singer, when applied to $K$-theory gives a perfectly good model.} Let $\mathrm{Struct}(X)$ be the monoid of \emph{structured Vector bundles} on $X$, that is vector bundles with  connection on $X$ modulo the relation $(V,\nabla)\sim (V',\nabla')$ iff $V$ is isomorphic to $V'$ and
\[CS[(V,\nabla),(V',\nabla')]\in\ud\Omega^\bullet(X),\]
where $CS$ is the Chern-Simons difference between the (isomorphic) vector bundles. Then $\diffGroup{K}^0(X)$ is the Grothendiek group generated by $\mathrm{Struct}(X)$: in other words one may intuitively think of elements of $\diffGroup{K}^0(X)$ as refining elements of $K^0(X)$ by endowing vector bundles with connection, and the map $c:\diffGroup{K}^0(X)\to K^0(X)$ is simply the forgetful map, forgetting the connections. The map $\Curv:K^0(X)\to\Forms^\even(X)$ is given by the Chern-Weil formula for the Chern character form of a vector bundle with connection: 
\[
\Curv:(V,\nabla)\mapsto \exp\left(-\frac{i}{2\pi}\nabla^2\right).
\]
Tensor product of structured vector bundles gives $\diffGroup{K}^0(X)$ a ring structure.
\end{example}
\subsubsection{Co-cycle models and gauge groups}
In formulating gauge theories one often encounters fields that transform under the gauge group, but in the end one divides out by the action of this group and ends up with gauge invariant quantities. The necessity of dealing with fields rather than just their gauge equivalence classes is easily seen when examining questions of \emph{locality}: fields may be glued together, but gauge equivalence classes cannot. In a similar manner, it is often important to have access to the co-cycles underlying the differential cohomology groups of interest. These have the crucial property that they may be glued. Just as with gauge-fields, differential co-cycles have symmetries: they form a (higher) groupoid. In fact, there are several classes of symmetries of a differential co-cycle that one can naturally consider -- in other words, there are several natural groupoids associated with the set of differential co-cycles. We will principally be concerned with two of these: here symmetries are ``geometric" -- the morphisms carry geometry.

The paradigmatic example to think about is the group $\diffGroup{H}^2(X)$. We recall that this classifies principal $U(1)$-bundles with connection, and a natural set of co-cycles is the set of principal $U(1)$-bundles with connection on $X$. There are several natural groupoids associated with this set. The obvious morphisms to consider are those that preserve the geometry: the connection preserving isomorphisms. Taking principal $U(1)$-bundles with connection modulo these returns precisely $\diffGroup{H}^2(X)$. There are, however, in this case, a fairly restrictive category of morphisms: the morphisms between two co-cycles form a torsor for locally constant $U(1)$-valued functions on $X$: $H^1(X;U(1))$. Another way to understand the groupoid is to think about the trivialisations: a geometry preserving trivialisation of a principal $U(1)$-bundle with connection is given by a choice of a flat section. Any other flat section is then obtained by rotating the chosen section on each component of $X$; in other words, by acting a locally constant $U(1)$-valued function $f:X\to U(1)$ on the chosen section. Thus the geometry-preserving trivialisations form a torsor for locally constant $U(1)$ valued functions on $X$.

One, may, however, wish to allow things to be a little flabbier, and consider the groupoid of $U(1)$-bundles with connection on $X$ with the morphisms being \emph{all} smooth isomorphisms. In this case the group of connected components is $H^2(X)$ -- in otherwords a topological invariant and rather small. However, the space of morphisms is in turn rather large. A $U(1)$-bundle with connection is now trivialised by \emph{any} smooth section, and these form a torsor for $C^\infty(X;U(1))$. However $C^\infty(X;U(1))=\diffEl{H}^1(X)$ -- the morphisms carry geometry!

This is a general phenomenon -- to any set of differential co-cycles one may associate two natural groupoids\footnote{In fact, there are four natural groupoids that one may associate to differential co-cycles \cite{Re}.}:
\begin{itemize}
\item The \emph{topological} groupoid: here the morphisms preserve the geometry and connected components recover the differential cohomology group, but the space of morphisms is restricted -- the space of trivialisations of an object is a torsor for $E^\bullet(X;\R/\Z)$,
\item and the \emph{geometric} groupoid. This has a lot more morphisms, so that connected components are naturally identified with $E^\bullet(X)$. However, the morphisms carry geometry: the space of trivialisations of a co-cycle is a torsor for $\diffGroup{E}^{\bullet-1}(X)$.
\end{itemize}
Gauge transformations of gauge fields are morphisms in the first groupoid, but very often one is interested in the second in applications, and that is the case for us. Whenever we talk about a ``trivialisation" of a co-cycle (or field), we shall allow the trivialisation to ``destroy the geometry", but the trivialisations will themselves carry geometry. 

A detailed discussion of these two groupoids in the case of ordinary differential cohomology is found in \cite{KaVa}.
\subsection{Twisted differential $K$-theory}\label{sec:twist}
As we shall see, \emph{twists} of differential $K$-theory will play a key roll in studying Ramond-Ramond fields. The subject of twisted differential $K$-theory is rather subtle, and the mathematics is still being developed. Hence we shall content ourselves to describe those mathematical features we need for the purposes of our work, and refer the interested reader to the review of Bunke and Schick (see Ch. 7 of \cite{BuSchiR})  and the references therein for further details.

Twisted differential $K$-theory is meant to be a geometric refinement of twisted $K$-theory in the same sense that differential $K$-theory is a geometric refinement of $K$-theory. Thus a logical pre-cursor to any discussion of twisted differential $K$-theory is a discussion of twists of $K$-theory, and in what sense they may be geometrically refined. The full range of twists of $K$-theory is somewhat intractible, so we shall focus our attention to an interesting subset of twists which we shall call the geometric twist. These twists arise from bundle gerbes (or alternatively $PU(H)$-bundles), and are classified by $H^3(X)$ for a manifold $X$. As we shall see, their interest for us is that they are the twists needed to discuss the influence of $B$-fields on the physics of Ramond-Ramond fields. 

To a given a bundle gerbe $B$ on a smooth manifold $X$, one may associate the twisted $K$-theory group $K^B(X)$. This is a module for $K(X)$, and an isomorphism of bundle gerbes $B\to B'$ induces an isomorphism of groups $K^B(X)\to K^{B'}(X)$. The twists arising from bundle gerbes are thus classified by $H^3(X)$. However, given one \emph{cannot} (canonically) twist $K$-theory by a class $[B]\in H^3(X)$ -- one needs an actual twisting object! Said differently: twists form a groupoid, not a set, and the morphisms are important.

Given that bundle gerbes twist $K$-theory, it is reasonable their geometric refinement, bundle gerbes with connection and curving, to twist differential $K$-theory, and this is indeed the case. Thus, given a bundle gerbe with the appropriate geometry, $\diffEl{B}$, one may form the group $\diffGroup{K}^{\diffEl{B}}(X)$, which is a module for $\diffGroup{K}(X)$. Again the \emph{category} of twists is important and gives rise to an key subtlety! Bundle gerbes with geometry are cocycles for $\diffGroup{H}^3(X)$, and as discussed in the previous section, there are several natural groupoids one may associate with these. The one appropriate to twisting $K$-theory is the ``geometric groupoid". Thus we are concerned with the groupoid of bundle gerbes with connection and curving, with morphisms being any smooth bundle gerbe isomorphisms (not just those preserving the geometry): any smooth isomorphism of bundle gerbes with connection and curving $\diffEl{B}\to\diffEl{B}'$ induces a homomorphism of groups $\diffGroup{K}^{\diffEl{B}}(X)\to\diffGroup{K}^{\diffEl{B}'}$. In fact, the twists of differential $K$-theory form a 2-groupoid, the groupoid of automorphisms of the trivial object being naturally identified with the groupoid of line bundles with connection on $X$. 

Just as with un-twisted differential $K$-theory there are various structure maps, giving rise to exact sequences
\begin{gather*}
\xymatrix{
0\ar[r]&\left(\frac{\Forms(X;V)^{\bullet-1+\Curv\diffEl{B}}}{\Curv\Forms(X;V)^{\bullet-2+\Curv\diffEl{B}}}\right)\ar[r]^<<<<<\formsmap&\diffGroup{K}^{\bullet+\diffEl{B}}(X)\ar[r]^c&
K^{\bullet+c(\diffEl{B})}(X)\ar[r]&
0
},
\\
\xymatrix{
0\ar[r]&K^{\bullet-1+c(\diffEl{B})}(X,\mathbb{R}/\mathbb{Z})\ar[r]&\diffGroup{K}^{\bullet+\diffEl{B}}(X)\ar[r]^>>>>>{\Curv}&\Forms_{\ud=0}(X;V)^{\bullet+\Curv\diffEl{B}}.
}
\end{gather*}
where $V=R[u,u^{-1}]$, with the degree of $u$ being two, and $\Forms(X,V)^{\bullet+\Curv\diffEl{B}}$ being the complex $(\Forms(X,V),\ud+H)$, where $H=\Curv\diffEl{B}\in\Forms^3(X)$ is the curvature of the bundle gerbe with connection and curving.
\subsubsection{Push forwards}\label{sec:push}
Just as in $K$-theory one may push-forward differential $K$-theory co-cycles. However, this push-forward not defined for any smooth map of spin-$\C$ manifolds -- one needs to give a geometric refinement the topological $K$-orientation (in other words the spin-$\C$ structure) as well as include some geometric data with the map. The appropriate geometric refinement of the spin-$\C$ structure is the introduction of a connection to the spin-$\C$ line bundle. The map needs to be refined to \emph{Riemannian} map. For a smooth submersion $\rho:X\to Y$ this amounts to choosing a metric along the vertical tangent bundle $T(X/Y)$ (essentially a Riemannian structure along the fibres) as well as a smooth projection $P:T(X)\to T(X/Y)$. One should think of a Riemannian map as defining a family of Riemannian manifolds, allowing one to do Riemannian geometry in families. In particular, the data then allows one to define a ``Levi-Civita" connection $\nabla^{X/Y}$ on $T(X/Y)$.

Given a geometrically $K$-oriented Riemannian map $f:X\to Y$ one may define a pushforward $f_*:\diffGroup{K}^\bullet(X)\to\diffGroup{K}^{\bullet+\dim X-\dim Y}(Y)$, as long as one has appropriate compactness. The pushforward refines the pushforward defined on $K$-theory and integration of differential forms. However, as a consequence of the index theorem, the push forward does not commute with the Chern character! Instead, one has a Riemann-Roch theorem~\cite{FrLo}:
\[
f_*\Ahat(\Omega^X)\Curv x = \Ahat(\Omega^Y)\Curv f_* x
\]
for $x\in\diffGroup{K}(X)$, and where the differential forms $\Ahat(\Omega^X)$ and $\Ahat(\Omega^Y)$ are constructed by evaluating the A-hat series on the curvature forms of the Levi-Civita connections on $X$ and $Y$ respectively.


It is reasonable to expect that one may refine the pushforward of twisted $K$-theory groups similarly to a pushforward of differential $K$-theory. However, as we have argued, the presence of the twist should modify the geometry! In particular, a cocycle $\diffEl{B}\in\diffGroup{H}^3(X)$ induces a generalised geometry, and so one obtains a modified Levi-Civita connection and the push-forwards should take this into account. We conjecture that this correction should take the following form.
\begin{theorem}\label{th:twistrr} Let $X\to T$ be a spin Riemannian family with compact fibres, and $\diffEl{B}\in\diffGroup{H}^3(X)$ be a differential 3-co-cycle on $X$. Then there is a bilinear pairing $\diffGroup{K}^{\diffEl{B}+\bullet}(X)\otimes\diffGroup{K}^{\diffEl{B}+\bullet}(X)\to\diffGroup{K}^{\bullet-\dim X/T}(T)$ defined by
\[
(\diffEl{x},\diffEl{y})=\int_{X/T}\diffEl{x}\cdot\theta(\diffEl{y})
\]
where $\theta:\diffGroup{K}^{\diffEl{B}+\bullet}(X)\to\diffGroup{K}^{-\diffEl{B}+\bullet}(X)$ is the smooth $-1$-Adams operation. Furthermore,
\[
\ch(\diffEl{x},\diffEl{y})=\int_{X/T}\hat{A}(\Omega^{X/T}_{\diffEl{B}})\mukai{\ch\diffEl{x}}{\ch \diffEl{y}},
\] 
where the pairing on ($\diffEl{B}$-twisted) differential forms is the Mukai pairing, and $\Omega^{X/T}_{\diffEl{B}}$ is the curvature of the torsionful generalised (relative) Levi-Civita connection determined by $\diffEl{B}$.
\end{theorem}
\section{Maxwell theories}\label{sec:maxwell}
In this section we summarise the framework for generalised abelian gauge theories first described in \cite{Fr:Dirac}. We will see that Dirac charge quantisation leads to a description of fields in these theories in terms of (generalised) \emph{differential cohomology}.

To begin we examine ordinary four dimensional Maxwell theory. Let $X$ be four-dimensional Minkowski spacetime, which we will assume to have a  space-time splitting $X=\R\times Y$, where $Y$ is a Riemannian 3-manifold. In classical Maxwell theory, all of the information is contained in the field-strength $F\in\Forms^2(X)$. In terms of this, the Maxwell equations are written
\begin{align*}
\ud F&=0,\\
\ud*F&=j_E,
\end{align*}
where $j_E\in \Forms^3(X)$ is the electric current density, a differential form compactly supported when restricted to spatial slices. The electric charge at time $t$ is given by
\[
q_E=\int_Y i_t^*j_E,
\]
where $i_t:Y\to X$ is the inclusion of $Y$ at time $t$. The electric charge may be any real number, and as a consequence of the field equations is conserved. 

It is natural to generalise the above situation: we allow space-time to have arbitrary dimension $X^{n+1}=\R\times Y^n$, and the field-strength to be a differential form $F\in\Forms^p(X)$ obeying equations of motion
\begin{align}
\ud F&=j_M,\label{eq:maxwellM}\\
\ud*F&=j_E;\label{eq:maxwellE}
\end{align}
where $j_E\in\Forms^{n-p+2}(X)$, $j_M\in\Forms^{p+1}(X)$, respectively the electric and magnetic current densities, are differential forms compactly supported in the spatial direction. The electric and magnetic current densities are closed differential forms as a consequence of the field equations, and hence, restricting them to a spatial slice, define classes in the compactly supported de Rham cohomology $q_E=[i_t^*j_E]\in H_c^{n-p+2}(Y,\R)$, $q_M=[i_t^*j_M]\in H_c^{p+1}(Y,\R)$, respectively the total electric and magnetic charge. The field equations also show the classes $q_E$ and $q_M$ are independent of the slice chosen. In the case of classical electromagnetism, where $Y=\R^3$, and $F\in\Forms^2(X)$, one sees that $q_E\in H^3(Y,\R)\cong\R$ is a real number obtained by integrating the current density over a spatial slice, recovering the usual notion of total electric charge.

The field equations (Eqs. \eqref{eq:maxwellM} and \eqref{eq:maxwellE}) imply that the classes in de Rham cohomology obtained by restricting $j_E$, $j_M$ to spatial slices vanishes (the field strength and its Hodge dual providing canonical trivialisations of these classes), so that
\begin{align*}
q_E&\in\ker\left(H^{n-p+2}_c(Y,\R)\to H^{n-p+2}(Y,\R)\right),\\
q_M&\in\ker\left(H^{p+1}_c(Y,\R)\to H^{p+1}(Y,\R)\right).
\end{align*}

We now examine how the picture changes when we quantise the theory. Dirac argued that charges must be quantised, that is that they lie on a lattice. It is natural to suppose that this lattice comes from the image of integer cohomology in real cohomology, i.e.~that 
\begin{align*}
q_E&\in H^{n-p+2}_c(Y,\Z)\hookrightarrow H^{n-p+2}_c(Y,\R),\\
q_M&\in H^{p+1}_c(Y,\Z)\hookrightarrow H^{p+1}_c(Y,\R),
\end{align*}
although we will see later that this is not the physically relevant possibility. Thus we see that electric and magnetic current densities should locally be like differential forms, but might globally contain information coming from integral cohomology classes, and it seems reasonable that they in fact be refined to \emph{differential} co-cycles\footnote{More precisely, objects in the the geometric groupoid  described in section \ref{sec:diffcoh}.} $\diffEl{j}_E\in\diffGroup{H}^{n-p+2}_c(Y)$, $\diffEl{j}_M\in\diffGroup{H}^{p+1}_c(Y)$. The electric and magnetic charges are then the restrictions of the characteristic class map to spatial slices, and the current densities being the ``curvatures'' of the respective co-cycles:
\begin{align*}
q_{E,M}&=i_t^*\cmap{\diffEl{j}_{E,M}},\\
j_{E,M}&=\Curv{\diffEl{j}_{E,M}}.
\end{align*}
One would like to imagine that the field strength might also be refined to a differential co-cycle, but Eq.~\eqref{eq:maxwellM} obstructs any naive sense of doing so. In order to simplify matters, we first examine the situation where there are no magnetic sources, so that the magnetic current density vanishes. In this case, the field-strength is closed, and we interpret it as the curvature of a differential cohomology co-cycle $\diffEl{A}\in\diffGroup{H}^p(X)$. In the case where the characteristic class vanishes (when $\diffEl{A}$ is \emph{topologically trivial}) refining $F$ to $\diffEl{A}$ amounts to the choice of an electro-magnetic potential $A\in\Forms^{p-1}(X)$. We define the fieldstrength map
$F:\diffGroup{H}^p\to\Forms^p(X)$ to be given by
\[
F:\diffEl{x}\mapsto\Curv{x}.
\]

One often imagines the electric and magnetic current densities as being associated to charged objects. For example, in electromagnetism, moving electrons are supposed to give rise to electric currents. This intuition is easily incorporated into the formalism. Electric charges are supported on submanifolds of dimension $p-1$, $i_E:W_E^{p-1}\to X$, thought of as $(p-2)$--dimensional objects moving through time, and magnetic charges are supported on manifolds of dimension $n-p$, $i_M:W_M^{n-p}\to X$. The induced electric or magnetic current is then the Poincar\'e dual of the included submanifold. Defining Poincar\'e duality in the world of differential cohomology either requires allowing distributional forms, or some choice of Thom form that rapidly decays away from the included submanifold. While these details are important for a proper mathematical treatment, they do not concern us here.
\begin{example}[Electromagnetism]
The paradigmatic example is electromagnetism. Here $n=3$, $p=2$. Thus $\diffEl{A}\in\diffGroup{H}^2(X)$ -- a principal $U(1)$-bundle with connection. The connection is the electromagnetic potential, and its fieldstrength is simply the curvature of the connection: in particular, we note that the fieldstrength is necessarily integral. The electric current density is a co-cycle in $\diffGroup{H}^{3}(X)$. As discussed briefly above, electrons moving through space-time trace out a one-dimensional submanifold $i_E:W_E\to X$ and the induced electric current is the Poincar\'e dual to this submanifold. Symbolically, $\diffEl{j}_E=(i_E)_*1$, where the pushforward is in differential cohomology. 
\end{example}
\begin{example}[$B$-fields]
An important example for us is that of $B$-fields. These are ``two-form" fields on ten-dimensional space time (i.e.~$p=3$, $n=9$) and thus $B$-fields are co-cycles in $\diffGroup{H}^{3}(X)$: they are bundle-gerbes with connection and connective structure. As we saw earlier, these are precisely the sort of objects that can be used to twist $K$-theory.
\end{example}
\begin{example}[Surfaces]
Our last example is the toy setting where $n=2$, $p=1$: in other words we're looking at scalar fields on a surface $X$. The field $\diffEl{A}$ is a $U(1)$ valued function. Electric charges are supported on points, and the induced electric current is an element of $\diffGroup{H}^2(X)$: a $U(1)$-bundle with connection. One may interpret the charges as being the divisor associated to the $U(1)$-bundle.
\end{example}
Allowing for magnetic charges changes the picture somewhat: the equation
\[
\ud F=j_M
\]
means that the fieldstrength is not closed, but instead trivialises the (necessarily exact) differential form $j_M$. Thus the fieldstrength can no longer be a curvature of a differential cocycle. Instead, the nature of the gauge field changes: it becomes a \emph{trivialisation} of the magnetic current. Thus
\[
\diffEl{A}:0\to\diffEl{j}_M.
\]
A simple example should illustrate the geometry here.
\begin{example}[Surfaces continued$\ldots$]
We return to the example of a scalar field on a surface. Magnetically charged objects are points on $X$, and the magnetic current is a $U(1)$-bundle with connection associated the divisor defined by these points: $\diffEl{j_M}\in\diffGroup{H}^2(X)$. The nature of the scalar field is forced to change: it is no longer a $U(1)$-valued function (i.e.~an element of $\diffGroup{H}^1(X)$) but instead is supposed to be a trivialisation
\[
\diffEl{A}:0\to\diffEl{j}_M;
\]
in other words, the field is now a section of the $U(1)$-bundle defined by the magnetic current. If we denote the $U(1)$-bundle with connection determined by the magnetic current by $(P_M,\nabla^M)\to X$, and the field by $\phi:X\to P_M$, then the fieldstrength of $\phi$ is given by $F(\phi)=\nabla^M\phi$, and thus $\ud F(\phi)=\Curv(\nabla^M)$ as expected.
\end{example}

Up until now we have assumed charges are quantised by the Eilenberg-McClane cohomology of a space, but there is a priori reason that this be so: one might imagine that charges are quantised by some generalised cohomology theory $E$, forcing the use of differential $E$-cohomology to describe the currents and fields of the theory. This, in fact, turns out to be crucial in the case of interest -- Ramond-Ramond fields and $D$-branes -- where some flavour of $K$-theory (depending on the precise flavour of String theory being considered) classifies $D$-brane charges. We will henceforth be in this setting: we assume spacetime to have dimension $n+1$, and assume the field to be in "degree $p$" so that $\diffEl{j}_E\in\diffGroup{E}^{n-p+2}_c(X)$, $\diffEl{A}:0\to\diffEl{j}_M\in\diffGroup{E}^{p+1}_c(X)$. We must also allow a slight generalisation of the definition of fieldstrength: the fieldstrength map is a map $F:\diffGroup{E}^\bullet\to\Forms(X;V)^\bullet$, where $V^\bullet=E(\pt)\otimes\R$ defined by 
\[
F:\diffEl{x}\to \omega_E\Curv(\diffEl{x})
\]
where $\omega_E$ is a closed, invertible differential form chosen as a part of the definition of the theory.
\subsection{The action}
Abelian gauge theories admit a Lagrangian formulation, and an action principle. In this section we examine the action in the light of our realisation that the dynamical variables in the gauge theories are differential co-cycles. For convenience, we work in Riemannian signature, and imagine that everything has been Wick rotated. As explained in the last paragraph of the previous subsection, we now assume charges are classified by some generalised cohomology theory $E$, and thus $\diffEl{j}_E\in\diffGroup{E}^{n-p+2}_c(X)$, $\diffEl{A}:0\to\diffEl{j}_M\in\diffGroup{E}^{p+1}_c(X)$. We recall the definition of the fieldstrength map $F:\diffGroup{E}^\bullet\to\Forms(X;V)^\bullet$, where $V^\bullet=E(\pt)\otimes\R$ defined by 
\[
F:\diffEl{x}\to \omega_E\Curv(\diffEl{x})
\]
where $\omega_E$ is a closed, invertible differential form chosen as a part of the definition of the theory. In order to write down the action, we need to restrict the choice of generalised cohomology: $E$ is postulated to have maps 
\begin{gather*}
E^1\to H^1,\\
E^2\to H^2,\\
\pi_{-\bullet}E_\R\to \pi_{-\bullet}H_\R\cong\R,
\end{gather*}
where by $E_\R$ (resp.~$H_\R$) we mean the $E$-theory (resp.~Eilenberg-McClane cohomology) with $\R$-coefficients. We are only interested in $E=H$, where the maps are obvious, and $E=K$, where the maps are respectively the ``determinant map" for the first two, and ``setting $u$ to zero" for the last (where we recall $K^\bullet(\pt)=\R[u,u^{-1}]$, with $u$ a formal variable in degree two. The existence of these maps gives rise to canonical maps
\[
\det:\diffGroup{E}^{1,2}\to\diffGroup{H}^{1,2}
\]
which we will soon need in defining the action.

Let us begin with the simplest case: the electric and magnetic charge densities vanish, and all that remains is the electromagnetic potential $\diffEl{A}\in\diffGroup{E}^p(X)$. The contribution to the action of the electromagnetic field $\diffEl{A}$ is just the usual
\[
S_{\diffEl{F}}=\frac{1}{e}\int_{X}\,F(\diffEl{A})\wedge*F(\diffEl{A}),
\] 
where $e$ is the electro-magnetic coupling, and we recall $F=\Curv\diffEl{F}$. This is clearly gauge-invariant, all the expressions in the integral being so.

Now we include the electric charge-density, $\diffEl{j}_E$. Classically, its contribution to the action is given by
\[
S_{j_E}``="(const)\int_X j_E\wedge A,
\] 
where $A$ is a choice the electromagnetic potential. However, this is not gauge invariant, but only the exponentiated action is. When refined to differential cohomology, it is this exponentiated action that we recover. We note that
\[
\diffEl{j}_E\cdot\diffEl{F}\in\diffGroup{E}^{n+1}(X)
\]
so that pushing forward to a point we have
\[
\int_X\diffEl{j}_E\cdot\diffEl{F}\in\diffGroup{E}^1(\pt).
\]
The exponentiated action is then defined to be
\[
e^{iS_{\diffEl{j}_E}}=\det\int_X\diffEl{j}_E\cdot\diffEl{F}\in \diffGroup{H}^1(\pt)=\R/\Z.
\]
It is often illuminating to think in families $X\to T$ (now of relative dimension $n$). The contribution to the action then becomes
\[
e^{iS_{\diffEl{j}_E}}=\det\int_{X/T}\diffEl{j}_E\cdot\diffEl{F}\in\diffGroup{H}^1(T),
\]
an $\R/\Z$ valued function of the parameter space $T$. To summarise, thus far the action reads
\begin{align*}
e^{iS}&=\exp{iS_{\diffGroup{F}}}\exp{iS_{\diffGroup{j}_E}}\\
&=\exp\left(i\frac{1}{e}\int_{X}\Curv{\diffEl{F}}\wedge*\Curv{\diffEl{F}}\right)\cdot\det\int_X\diffEl{j}_E\cdot\diffEl{F}.
\end{align*}
We now wish to include the contribution of the magnetic charge density to the discussion. It will be illuminating to work in families. Recall that introducing the magnetic charge density changes the nature of the electromagnetic field: instead of being a differential cohomology co-cycle, it now is a trivialisation of the magnetic charge density (this being the content of the equation $\ud F=j_M$). Thus the contribution to the exponentiated action 
\[
e^{iS_{\diffEl{j}_E}}=\det\int_{X/T}\diffEl{j}_E\cdot\diffEl{F}
\]
is no longer a co-cycle in $\diffGroup{H}^{1}(T)$, but rather a trivialisation of 
\[
\det\int_{X/T}\diffEl{j}_E\cdot\diffEl{j}_M\in\diffGroup{H}^{2}(T),
\] 
i.e.~a section of the $U(1)$ bundle $\det\int_{X/T}\diffEl{j}_E\cdot\diffEl{j}_M$. This line bundle may be \emph{anomalous}; in other words, its class in $\diffGroup{H}^2(T)$ may non-zero (although the fact that it has a section does say its characteristic class vanishes: the anomaly is necessarily local). Even if the anomaly vanishes, one needs an explicit choice of trivialisation (in the \emph{topological groupoid}) in order to interpret the section  this section as an $\R/\Z$ valued function requires an explicit trivialisation of this line bundle -- a choice of a flat section of this bundle. 

Putting everything together, we see that the general (exponentiated) action is given by
\[
e^{iS_{\diffEl{F}}}e^{iS_{\diffEl{j}_E}}:0\to\det\int_{X/T}\diffEl{j}_E\cdot\diffEl{j}_M,
\] 
and may only be interpreted as an $\R/\Z$ valued function on $T$ upon a choice of trivialisation of $\det\int_{X/T}\diffEl{j}_E\cdot\diffEl{j}_M$.
\subsection{Self-duality}
Classically, the electromagnetic field is said to be self-dual when it obeys the additional constraint $F=*F$. We now discuss self-duality in the context of generalised abelian gauge theories, and will see that self-duality imposes additional subtleties on charge quantisation. 

We will approach self-duality in the Wick-rotated setting, and assume now that $X$ is a Riemannian manifold, with $\dim X=n$. Let us suppose further that the charges in the theory are quantised by a generalised cohomology theory $E$. By the discussion in the previous section, the electric charge density is then a co-cycle $\diffEl{j}_E\in\diffGroup{E}^{n-p+1}(X)$, the magnetic charge density a co-cycle $\diffEl{j}_M\in\diffGroup{E}^{p+1}(X)$, and the gauge field a \emph{topological} trivialisation $\diffEl{F}:0\to\diffEl{j}_M$. 

We now wish to impose a \emph{self-duality constraint}. Following Freed and others \cite{Fr:Dirac} we define this to be:
\begin{itemize}
\item an automorphism $\theta:\diffGroup{E}^\bullet\to\diffGroup{E}^{n-\bullet+2}$ with the property that the bilinear form on pairs of degree $d+1$ co-cycles defined by
\[
(\diffEl{x},\diffEl{y})=\int_{X/T}\diffEl{x}\cdot\theta(\diffEl{y})
\]
is symmetric for any fibre bundle $X\to T$ of fibre dimension $d$.
\item For each fibre bundle as above, a quadratic map $q_{X/Y}:\diffGroup{Z}^{d+1}_{E}(\mathcal{X})\to\diffGroup{Z}^2_{E}(T)$ refining the above bilinear pairing.
\end{itemize}
Both the bilinear pairing and the quadratic refinement should be natural, in the sense spelled out in \cite{Fr:Dirac} In particular, a trivialisation of $\diffEl{x}\in\diffGroup{Z}^{d+1}_E(\mathcal{X})$ should give  a canonical trivialisation of $q_{\mathcal{X}/Y}(\diffEl{x})$. While the choice of a quadratic refinement of the bilinear form is crucial when discussing subtle questions of quantisation \cite{HoSi, FrMoSe}, or our purpose we can ignore it and work entirely in terms of the bilinear form. This has the advantage of making the presentation clearer, and, in particular, will illuminate a crucial point in our treatment of Ramond-Ramond fields coupling to $D$-Branes in the presence of $B$-fields.

The self-duality constraint may be formulated very simply from the discussion above. It is that
\[
\diffEl{j}=\theta(\diffEl{j}_E).
\]
We will thus henceforth only refer to the magnetic current $\diffEl{j}$, interpreting the gauge field as a trivialisation $\diffEl{A}:0\to\diffEl{j}$.

The kinetic term in the self-dual action remains the same, but self-duality gives rise to a square-root in the contribution to the action arising from the coupling between the gauge field and the current. Recall from previous discussion that without self-duality, the coupling would be
\[
e^{iS_{\diffEl{j}}}=\det\int_{X/T}\theta(\diffEl{j})\cdot\diffEl{A}=\det(\theta(\diffEl{j}),\diffEl{A}):0\to\det(\theta(\diffEl{j}),\diffEl{j}).
\]
In the presence of self-duality, the contribution to the action should be $\exp[iS_{\diffEl{j}}]^{1/2}$. Taking this square root is precisely what the choice of the quadratic form $q_{X/T}$ does, and one defines the contribution to the action to be
\[
\exp[iS_{\diffEl{j}}]^{1/2}=\det q_{X/T}(\diffEl{A}):0\to\det q_{X/T}(\diffEl{j}).
\]
We will however continue working with the ``un-square-rooted" action $e^{iS_{\diffEl{j}}}=\det(\theta(\diffEl{j}),\diffEl{A})$: the subtlety of the choice of quadratic refinement does not affect our discussion and obscures our main point.
\subsection{Ramond-Ramond fields}
We now apply the framework reviewed in the previous section to Ramond-Ramond fields in type II, setting $B$ to zero. In the next section we turn to the case of main interest to the paper, $B\ne0$. We will continue to work in a Wick-rotated setting, and take $X$ to be a 10-dimensional compact Riemannian spin manifold. $D$-branes carry Ramond-Ramond charges, and are submanifolds $i:W\hookrightarrow X$ of appropriate dimension (odd, resp.~even for type IIA/B) with vector bundles $(V,\nabla)\to W$ (the Chan-Paton bundle). In \cite{Witten1998, MiMo} it was realised that $D$-brane charges are quantised by an appropriate version of $K$-theory: $K^0$ for type IIB, $K^1$ for type IIA, which is twisted in the presence of $B$-fields.
\subsection{$B=0$, no $D$-branes}
We begin with the simplest situation: no $D$-branes, and zero $B$-field. Because $K$-theory carries Ramond-Ramond charges, Ramond-Ramond fields are co-cycles in $\diffGroup{K}^{0}(X)$ (resp.~$\diffGroup{K}^1(X)$) in type IIB (resp.~type IIA). The field-strength map is normalised so that, for $\diffEl{C}\in\diffGroup{K}^{\bullet}(X)$,
\[
\diffEl{F}(\diffEl{C})=\sqrt{\hat{A}(X)}\ch(\diffEl{C}),
\]
where $\hat{A}(X)$ is the $\hat{A}$-form formed from the Levi-Civita connection on $X$ (or in the case of a Riemannian family, $\hat{A}(\Omega^{X/T})$)\footnote{There are various ways to justify the inclusion of the $\hat{A}$-factor. Perhaps the most convincing is that one wants the bilinear pairing introduced soon to commute with fieldstrength map and pairing differential forms. Push-forward and the Chern character commute up to a factor of $\hat{A}$, so that introducing the square-root of $\hat{A}$ in the definition of the fieldstrength map ensures this.}. Ramond-Ramond fields are self-dual: the map $\theta:\diffGroup{K}^{p}(X)\to\diffGroup{K}^{12-p}(X)$ is given by
\[
\theta:\diffEl{x}\mapsto u^{6-p}\bar{\diffEl{x}},
\] 
where $u\in K^{2}(\pt)$ is the inverse Bott element, and $\bar{\diffEl{x}}$ is induced by complex conjugation of vector bundles with connection: said in more sophisticated language, theta is a differential refinement of the $-1$ Adams operation on $K$-theory. The bilinear pairing $\diffGroup{K}^p(X)\otimes\diffGroup{K}^p(X)\to\diffGroup{H}^2(T)$ for a spin and Riemannian family $X/T$ is given by
\[
(\diffEl{x},\diffEl{y})=\det\int_{X/T}\theta(\diffEl{x})\cdot\diffEl{y}.
\]
At the level of fieldstrengths, an index-theory calculation shows that this is implemented by pairing the fieldstrengths using the Mukai pairing:
\begin{align*}
\ch(\diffEl{x},\diffEl{y})&=\left[\ch\int_{X/T}\theta(\diffEl{x})\cdot\diffEl{y}\right]_{(2),u=0}\\
&=\left[\int_{X/T}\hat{A}(X/T)\ch\theta(\diffEl{x})\cdot\diffEl{y}\right]_{(2),u=0}\\
&=\left[\int_{X/T}\hat{A}(X/T)\ch\theta(\diffEl{x})\wedge\ch\diffEl{y}\right]_{(2),u=0}\\
&=\left[\int_{X/T}F(\theta(\diffEl{x})\wedge F(\theta(\diffEl{y}))\right]_{(2),u=0}\\
&=\langle F(\diffEl{x}), F(\diffEl{y})\rangle_{(2),u=0}.
\end{align*}
\subsection{$D$-brane couplings when $B=0$}
We now incorporate $D$-branes into the picture. By self-duality, including them into the picture will immediately induce both electric \emph{and} magnetic currents, and so change the nature of the Ramond-Ramond field itself. As a first approximation, the $D$-brane will be a co-dimension $r$ submanifold ($r$ odd or even depending on whether we are in type IIA or B) $i:W\to X$ along with a vector bundle and connection $(V,\nabla)\to W$, the Chan-Paton bundle. Let $\diffEl{q}_V$ be the class in $\diffEl{K}^0(W)$ induced by $(V,\nabla)$. The magnetic current is then defined as 
\[
\diffEl{j}=u^{\left\lfloor\frac{r+p}{2}\right\rfloor-4}i_*(V,\nabla).
\]
There are two obstructions to the above definition of the magnetic current: in order to define the pushforward $i_*\diffGroup{K}^{0}(W)\to \diffGroup{K}^{10-r}(X)$, $W$ needs to be \emph{oriented} and its normal bundle needs to be \emph{spin}. We will henceforth assume the map $i$ to be oriented, but not that the normal bundle be spin. The pushforward is then defined from a \emph{twisted} differential $K$-theory on $W$: $i_*:\diffGroup{K}^{0-\diffEl{w}_2(\nu)}(W)\to\diffGroup{K}^{10-r}(X)$, where $\diffEl{w}_2(\nu)$ is a (differential) characteristic class induced by the second Stiefel-Whitney class of the normal bundle $\nu$. Thus the Chan-Paton bundle is no longer a simple vector bundle, but twisted by $\diffEl{w}_2(\nu)$. For example, in rank one, it is a $\text{spin}^c$-connection on $\nu$.

Self-duality implies that the electric current density is obtained from the magnetic current density, being given by
\[
\theta(\diffEl{j})=u^{1+\left\lfloor\frac{r-p}{2}\right\rfloor}i_*\bar(V,\nabla).
\]
The nature of the Ramond-Ramond field also changes. It is now not a co-cycle in $\diffGroup{K}^{\bullet}(X)$, but rather a (topological) trivialisation
\[
\diffEl{C}:0\to\diffEl{j}.
\]
The electric coupling term will now be a section of a line bundle, determined by (a square root of)
\[
(\theta(\diffEl{j}),\diffEl{j})=\det\int_{X/T}u^{r-4}i_*\bar{\diffEl{q}}_V\cdot i_*\diffEl{q}_V.
\]
Taking the square root (which is what the quadratic refinement does) may give rise to an anomaly, which Freed and Hopkins \cite{FrHo} argue cancels the anomaly arising from the fermions on $W$. The electric coupling is then a trivialisation of this line bundle (i.e.~a section), and is determined by 
\[
(\theta(\diffEl{j}),\diffEl{C})=\det\int_{X/T}u^{1+\left\lfloor\frac{r-p}{2}\right\rfloor}i_*\bar{\diffEl{q}}_V\cdot\diffEl{C}.
\]
Let us now write this in a form more familiar to the physics literature. We assume the $D$-brane has a Chan-Paton bundle, a complex vector bundle $(V,\nabla_V)\to W$, and that $W$ is $\text{spin}^c$ and the curvature of the $\text{spin}^c$ connection is $-2\pi i\eta\in\Forms^2(W)$. Let us now suppose that the Ramond-Ramond field is determined by a differential form $C$, so that its fieldstrength is $2\pi\sqrt{\hat{A}(X/T)}\ud C$. Then we may write the electromagnetic coupling as
\begin{align*}
(\theta(\diffEl{j}),\diffEl{C})&=\exp-2\pi i \left[\int_{X/T}\mukai{\diffEl{F}(i_*\diffEl{q}_V)}{C}\right]_{(0),u=0}\\
&=\exp-2\pi i \left[\int_{X/T}\mukai{\sqrt{\hat{A}(X/T)}\ch i_*\diffEl{q}_V}{C}\right]_{(0),u=0}\\
&=\exp-2\pi i \left[\int_{W/T}\mukai{\hat{A}(\nu)^{-1}\wedge e^{\eta/2}\wedge\sqrt{\hat{A}(X/T)}\ch{\nabla_V}}{i^*{C}}\right]_{(0),u=0}\\
&``=''\exp-2\pi i \left[\int_{W/T}\sqrt{\frac{\hat{A}(W/T)}{\hat{A}(\nu)}}e^{\eta/2}\mukai{i^*{C}}{\ch{\nabla_V}}\right]_{(0),u=0},
\end{align*}
where the second last line follows from Riemann-Roch. The last line would follow \emph{if the normal and tangent bundles of $W$ split geometrically}, in other words if $\hat{A}(X/T)=\hat{A}(W/T)\hat{A}(\nu)$. However, this is usually only the case at the level of characteristic classes, and \emph{not} the case at the level of forms (and one in general thus expects exact ``cross terms" involving derivatives of the metric in both tangent and normal directions).
\section{Incorporating $B$-field}\label{sec:rrfield}
We are finally at the point where we can incorporate $B$-fields into the picture. For us, a $B$-field will be taken to be a co-cycle $\diffEl{B}\in\diffGroup{H}^3(X)$: one may think of it as a bundle gerbe with connection and curving. As discussed in Section \ref{sec:gengeom}, the $B$-field determines a generalised geometry on $X/T$, and in particular, when $X/T$ is a Riemannian fibration, one obtains a vertical ``Levi-Civita" connection $\nabla^B$ on $T(X/B)$ (and hence, when the fibres are spin, also on the vertical spinors). 

The $B$-field also twists the differential $K$-theory of $X$, and our Ramond-Ramond fields, and the magnetic flux will be co-cycles in $\diffGroup{K}^{\bullet+\diffEl{B}}(X)$. We recall that the $-1$-Adams operation induces a canonical morphism from $\alpha^{-1}:\diffGroup{K}^{\bullet+\diffEl{B}}(X)\to \diffGroup{K}^{\bullet-\diffEl{B}}(X)$, and the morphism $\theta$ we used to define self-duality in the setting of Ramond-Ramond fields now becomes a morphism $\theta:\diffGroup{K}^{\bullet+\diffEl{B}}(X)\to\diffGroup{K}^{12-\bullet-\diffEl{B}}$ defined by
\[
\theta=u^{6-p}\alpha^{-1}.
\]
This now allows us to define a pairing $(\cdot,\cdot)_{\diffEl{B}}:\diffGroup{K}^{\bullet+\diffEl{B}}(X)\otimes\diffGroup{K}^{\bullet+\diffEl{B}}(X)\to H^{2}(B)$ by
\begin{equation}\label{eq:bpairing}
(\cdot,\cdot)_{\diffEl{B}}:\diffEl{x},\diffEl{y}\mapsto\det\int_{X/B,\diffEl{B}}\theta(\diffEl{x})\cdot\diffEl{y}
\end{equation}
which, as before, may be refined to a quadratic form.

This brings us to our first key point: the geometry used to define the pushforward in Eq.~\eqref{eq:bpairing} is the \emph{generalised} geometry determined by $\diffEl{B}$. In particular, we use the \emph{generalised} Levi-Civita connection $\nabla^{B}$ to form our Dirac operators, and the appropriate index theorem here is the local index theorem proved by Bismut for torsionful connections \cite{BisTor}:
\[
\lim_{t\to0}\mathrm{Tr}e^{-t\Dirac(\nabla^{\diffEl{B}})^2}=[\hat{A}(\nabla^{-B})]_{(10)},
\]
and our fieldstrength map needs to reflect this. The appropriate definition of the fieldstrength map $F:\diffGroup{K}^{\bullet+\diffEl{B}}(X)\to\Forms^{\bullet+\diffEl{B}}(X)$ in this context is thus given by
\[
F:\diffEl{x}\mapsto\sqrt{\hat{A}(\nabla^{-B})}\ch{\diffEl{x}}.
\]

A consequence of the appearance of the $B$-twisted geometry is that the pairing \eqref{eq:bpairing} \emph{is no longer symmetric} and thus must explicitly be symmetrised. We thus define the symmetrised bilinear pairing:
\begin{align*}
(\cdot,\cdot):\diffEl{x},\diffEl{y}&\mapsto\frac{1}{2}\left((\diffEl{x},\diffEl{y})_{\diffEl{B}}+(\theta(\diffEl{y}),\theta(\diffEl{x}))_{-\diffEl{B}}\right)\\
&=\det\left(\frac{1}{2}\int_{X/B,\diffEl{B}}\theta(\diffEl{x})\cdot\diffEl{y}+\frac{1}{2}\int_{X/B,-\diffEl{B}}\diffEl{y}\cdot\theta(\diffEl{x})\right).
\end{align*}

In the presence of $B$-fields, charges are quantised by $B$-twisted $K$-theory. Thus Ramond-Ramond fields become co-cycles $\diffEl{C}\in\diffGroup{K}^{p+\diffEl{B}}(X)$. More importantly, this imposes a restriction on admissible $D$-branes: a $D$-brane now is a submanifold $i:W\to X$ of dimension $r$ with a Chan-Paton bundle $(V,\nabla)\to W$ along with a \emph{choice of trivialisation} $\tau_B:0\to i^*\diffEl{B}\in\diffGroup{H}^3(W)$. This trivialisation induces canonical morphisms $\tau_B:\diffGroup{K}^{\bullet}(W)\to\diffGroup{K}^{\bullet+i*\diffEl{B}}(W)$, and $e^{B}:\Forms^{\bullet}(W)\to\Forms^{\bullet+\diffEl{B}}(W)$. Using the first morphism, we define the induced magnetic flux as follows
\[
\diffEl{j}=u^{\left\lfloor\frac{r+p}{2}\right\rfloor-4}i_*\tau_{\diffEl{B}}(V,\nabla).
\]
Following the computation in the previous section, we now see that the Ramond-Ramond field/$D$-brane coupling is computed on the bulk by
\begin{equation}\label{eq:dbranebulk}
(\theta(\diffEl{j}),\diffEl{C})=\exp-2\pi i \left[\int_{X/T}\frac{1}{2}\left(\mukai{\sqrt{\hat{A}(\Omega^{X/T}_{B})}\ch i_*\diffEl{q}_V}{C}+\mukai{\bar{C}}{\sqrt{\hat{A}(\Omega^{X/T}_{-B})}\ch i_*\diffEl{q}_{\bar{V}}}\right)\right]_{(0),u=0}.
\end{equation}
Pulling this back to the $D$-brane, the expression becomes complicated. In the next section, we will examine the integrand carefully, and for now content ourselves with showing the most important terms:
\begin{multline}
\label{eq:dbranecoupling}
(\theta(\diffEl{j}),\diffEl{C})\\
=\exp-2\pi i \bigg[\int_{W/T}\frac{1}{2}\mukai{u^{\left\lfloor\frac{r+p}{2}\right\rfloor-4}\hat{A}(\nu)^{-1}\sqrt{\hat{A}(\nabla^{X/T,\diffEl{B}})}e^{\eta/2}e^{B}\ch{\nabla_V}}{i^*{C}} \\
 +\int_{W/T}\frac{1}{2}\mukai{i^*{C}}{u^{\left\lfloor\frac{r+p}{2}\right\rfloor-4}\hat{A}(\nu)^{-1}\sqrt{\hat{A}(\nabla^{X/T,\diffEl{-B}})}e^{\eta/2}e^{B}\ch{\nabla_V}}\bigg]_{(0),u=0}+\cdots
\end{multline}
There are several key points to note:
\begin{itemize}
\item The connections used in the $\hat{A}$ expressions contain the $B$-field, leading to new terms in the derivative of $H$.
\item These occur symmetrically because of the two terms in the expression.
\item One does not expect $\hat{A}(\nabla^{X/T,\diffEl{\pm B}})$ to split into a tangent and normal part, and thus in general expects ``mixed derivative" terms to appear.
\end{itemize}

\section{$T$-duality, $B$-field and $D$-branes}\label{sec:tdual}

Let us start by observing that the ``leading" coupling on the $D$-brane bulk \eqref {eq:dbranecoupling} is of standard type and contains wedge products of forms pulled back from the bulk with the bulk quantitates. Indeed ignoring the symmetrisation in $B$, this term can  be written schematically  as 
\begin{equation}\label{eq:standard}
\int_{W/T} \left[i^*C \wedge \left(X(\nabla^{W/T, \pm B}, \nu) \wedge  \ch{ \nabla_V} \right) \right]_{p+1} \, ,
\end{equation}
where $C$ is the RR polyform and $p+1$ is the dimension of the $D$-brane bulk, and appears to be only a mild modification of  the $D$-brane couplings for $B=0$ (a shift of $\sqrt{\hat{A}(\nabla^{W/T,\diffEl{B}})}$ by an exact form). It has been argued recently \cite{BeckRob, Ga, GaM}, that such couplings cannot be invariant under $T$-duality, regardless the details of the form $X$. Moreover, bulk calculations (involving disk amplitudes with insertions of one RR and two NS vertex operators) indicate that such couplings are part of more general patter where the integrand is again a $(p+1)$-form, but the couplings  now involves  $n$ contractions between the  $(r+n)$-form $i^*C$ and  $s+n$-form $X_{(s,n)}(\nabla^{W/T, B}, \nu)$:\footnote{We have introduced  double-index notation, where the first index denotes the rank of the differential form, and second index denotes the number of contracted indices.} 
\begin{multline}
\label{eq:contra}
(\theta(\diffEl{j}),\diffEl{C}) =  \exp \sum_{r+s + q = p+1 }  \sum_n \\ 
-2\pi i \bigg[\int_{W/T}\frac{1}{2}\mukai{u^{\left\lfloor\frac{r+p}{2}\right\rfloor-4} X_{(s,n)}  (\nabla^{W/T, B}, \nu) \wedge \left( e^{B}\ch{\nabla_V} \right)_{(q,0)}  }{i^*{C_{(r,n)}}} \\
+  (-1)^n \int_{W/T}\frac{1}{2}\mukai{i^*{C_{(r,n)}}}{u^{\left\lfloor\frac{r+p}{2}\right\rfloor-4}X_{(s,n)}  (\nabla^{W/T, - B}, \nu) \wedge \left(e^{B}\ch{\nabla_V} \right)_{(q,0)}  }\bigg]_{(0),u=0}
\end{multline}
The precise form of $X_{(s,n)} (\nabla^{W/T, B}, \nu) $ depends on values of the differential rank  $s$ and the number of contractions $n$.  A particular feature of \eqref{eq:contra} is that  the terms with even/odd numbers of contractions $n$  contain only even/odd powers of $B$ or rather the curvature three-form $H$.  Note that the contraction does not affect $\ch{ \nabla_V}$. Shortly we shall simply take the Chan-Paton bundle to be trivial ignore this factor altogether.

In this section we shall argue that  the modified bulk $D$-brane couplings (\eqref{eq:dbranebulk}) are  $T$-duality invariant and reduce to form  \eqref{eq:contra} when restricted to the $D$-brane worldvolume. We shall not present any formal proofs or aim at being very general. Instead we shall just discuss in detail the simplest nontrivial illustration.

\subsection{Pure spinors, Mukai product and $T$-duality}
Consider a single $T$-duality along the isometry generated by a vector $v= \partial/ \partial t$ (the dual one form is given by $\imath_v e = 1$ and can be written as $e= \dd t +  a$. Without loss of generality one can consider $B$-field such that $\calL_v B=0$. Then the $T$-duality on the pure spinor $C$ is simply:
$$T_v C = \imath_v C + \dd t \wedge C $$
(see the discussion and the general case for $\calL_v B \neq 0$ in sec. 3.1 of \cite{GMPW}).

It is not hard to check that for $B = B_2 + b \wedge e$ and $C^{(-)} = e^{-B} \hat{C} = C_p +  {C} _{p-1} \wedge e = e^{-B_2}[ \hat{C}_p + ( \hat{C}_p \wedge b +  \hat{C}_{p-1} )\wedge e]$,
$$T_v C =  e^{-\tilde{B}} [ (\dd t + b) \wedge  \hat{C}_p  + \hat{C}_{p-1}] \, ,$$
where $\tilde{B} = (B_2 + b \wedge a) + a \wedge (\dd t + b)$.   By taking $B \rightarrow -B$ we could also define $C^{(+)}$, which will transform the same way under $T$-duality. 

Note that $T$-duality swaps $a$ and $b$  (topologically speaking the first Chern class of the circle bundle $c_1$ with $\int \imath_v H$) and sends $B_2 \, \mapsto \, B_2 + b \wedge a$. The latter corresponds to leaving horizontal component $H_3$ of the $H$-flux ($H= H_3 + H_2 \wedge e$) invariant under $T$-duality. 

One can check now that the Mukai paring of two spinors $ C^{(\pm)} $ and $\alpha^{(\pm)}$, which transform under $T$-duality in the above manner, is invariant under $T$-duality: $\langle C^{(\pm)},  \alpha^{(\pm)} \rangle =  \pm \langle T_v C^{(\pm)},  T_v \alpha^{(\pm)} \rangle$ . It is crucial here, that the signs of the $B$-field are correlated here and that the Mukai product is effectively ``no-$B$" ($\langle C^{(\pm)},  \alpha^{(\pm)} \rangle = \langle  \hat{C} ,   \hat{\alpha}  \rangle$) and the local one-forms $a$ and $b$ drop out. Hence the bulk $D$-brane couplings are invariant under $T$-duality provided $\alpha^{(\pm)}$ transforms as a pure spinor. We shall now turn to the discussion of this object and its $T$-duality properties . 

Before doing so, let us quickly comment on RR gauge invariance under $C \mapsto C + \dd \Lambda$. Clearly $\langle C,  \alpha \rangle$ is an invariant coupling since $\dd \alpha = 0$. Of course, this invariance persists  in presence of the isometry, but we would simply like to observe that it now requires cancellation of contributions from different parts of the poly-forms $C$ and $\alpha$.  Consider now a U(1)-fibered background $\pi: X \longrightarrow Y$  (with a the curvature of the principal U(1) bundle $\dd e = \pi^*F$   (locally $F = \dd a$)\footnote{Throughout this section, we shall be dropping the pull-backs $\pi^*$ in order not to clutter the formulae too much.}).  Now $C_{p-1} \mapsto C_{p-1} + \dd \Lambda_{p-2}$ while $C_p \mapsto C_p + \dd \Lambda_{p_1} + (-)^p \Lambda_{p-2} \wedge F$ , and (using integration by parts)
$$
\langle C,  \alpha \rangle \mapsto \langle C,  \alpha \rangle \pm  \langle  \Lambda_{p-2}  \wedge e,  (\dd \alpha_{10-p} - (-)^p \alpha_{9-p} \wedge F) \rangle \, .
$$
The latter is the vanishing of the horizontal component of $\dd \alpha$. When $T$-dualising a particular brane configuration, we should bare in mind that depending of the gauge invariance will require contributions from different $D$-branes to cancel out. One of the advantages of writing the $D$-brane couplings in the bulk rather that on brane world-volumes is that this invariance is less obscure.

As we have argued  the $D$-brane couplings are given by
\begin{equation}
\label{Dbra-co}
I_{\tiny \mbox{$D$-brane}}  = (\theta(\diffEl{j}),\diffEl{C})=\exp-2\pi i \left[\int_{X/T}\frac{1}{2}\left(\mukai{\alpha^+}{C}+\mukai{\bar{C}}{\alpha^-}\right)\right]
\end{equation}
and 
$$
\alpha^{(\pm)} = e^{\pm B} \sqrt{\hat{A}(\Omega^{X/T}_{B})}\ch i_*\diffEl{q}_V  \, .
$$
We are now ready to see that this form of the coupling ensures the $T$-duality invariance of $D$-brane actions but can account for the couplings \eqref{eq:contra} for $n >0$.

\subsection{$T$-duality invariance of $D$-brane couplings}\label{TDc}
We shall consider the simplest nontrivial case of relevant $D$-brane couplings  $I_{\tiny \mbox{$D$-brane}} $  for a D5-brane with a trivial Chan-Paton bundle  with only two-form RR field $C_2$ turned on. 
Then  $\alpha^{(\pm)} = X^{(\pm)}_4 \wedge \eta(W_6 \hookrightarrow M_{10}) $, and $X^{(\pm)}_4 =  \frac{1}{2} p_1^{(\pm)} (X) $ are simply Pontrjagin classes computed with a connection with torsion, i.e. differ form $\tr R^2$ by exact terms. 
$$
I_{\tiny \mbox{D5}} = \frac{1}{4} \, C_2 \wedge \left( p_1^+ (X) + p_1^- (X) \right) \wedge \eta  = C_2 \wedge X \wedge \eta
$$ 
On a U(1)-fibered background $\pi: X \longrightarrow Y$, $\eta(W_6 \hookrightarrow M_{10}) = \eta_4 + \tilde{\eta}_3 \wedge e$. After $T$-duality $\tilde{\eta}_3$ should be the Poincar\'e dual to $D6$ while $\tilde{\eta}_5 = \eta_4 \wedge (\dd t + ...)$ - to $D4$. 

Similarly, the four-form $X = X_4 + X_3 \wedge e$, and $\dd X_4 = X_3 \wedge F$ and $\dd X_3 = 0$. Moreover one can show $X_3 = \dd X_2$ for a globally defined $X_2$.\footnote{We are considering $X$ to be a four-form for sake of concreteness. For large enough dimensions of the worldvolume eight-forms built from $p_2^{(\pm)}$ and $(p_1^2)^{(\pm)}$ may appear as well. With obvious change in respective sub-scripts, all the statements concerning $T$-duality apply as well.} It may be convenient to write
$$
X = (X_4- F\wedge X_2) + \dd (X_2 \wedge e) \, .
$$
Now the two parts are separately closed, and have nice $T$-duality properties which will be important.\footnote{Derivations and explicit expressions can be found in \cite{LiuM}.}  In particular, $(X_4- F\wedge X_2)$ is $T$-duality invariant. $X_2 = \dd^{-1} (\imath_v X) = \dd^{-1} \left(\imath_v (X^{(+)} + X^{(-)})/2 \right)$ is not invariant but  maps under $T$-duality to $\tilde{X}_2 =  \dd^{-1} \left(\imath_v (X^{(+)} - X^{(-)})/2 \right)$. In other words, while the first part of $X$ is the same when written in terms of original and $T$-dual fields (the connection and the $B$-field) , the second part has an important flip of relative signs between $ \dd^{-1} (\imath_v X^{(+)})$ and $ \dd^{-1} (\imath_v X^{(-)})$ when passing from original fields to the dual ones. Notably if the original expression has only even powers of the $B$-field, we get odd powers of $\tilde{B}$ in the dual picture. Another way of saying all this is $X_2(T, \imath_v H) =   \tilde{X}_2 (\imath_{v} \tilde{H}, \tilde{T})$.

Returning to $I_{\tiny \mbox{D5}}$ we see that under $T$-duality $C_2 = c_2 + \tilde{c}_1 \wedge e   \longrightarrow [c_2 + \tilde{c}_1 \wedge a] \wedge \dd t - \tilde{c}_1$. The $T$-dual of $\alpha$ is equally un-pretty, and
\begin{eqnarray}
T_v  (I_{\tiny \mbox{D5}} ) &=&  - c_2 \wedge \dd t \wedge X_4 \wedge \tilde{\eta}_3 - [ \tilde{c}_1 \wedge X_4  + c_2 \wedge X_3 ] \wedge \eta \wedge \dd t \cr
&=&  - c_2 \wedge \tilde{e}  \wedge X_4 \wedge \tilde{\eta}_3 - [ \tilde{c}_1 \wedge X_4  + c_2 \wedge X_3 ] \wedge \eta \wedge\tilde{e}  \nonumber
\end{eqnarray}
where $\tilde{e} = (\dd t + b)$ has been introduced.

As expected we can see  $D6$ and $D4$ couplings. Bearing in mind that the RR gauge invariance may require relative cancellation between the branes, we may look at them separately. The $D6$ part holds no surprises and its coupling to RR three-form is of standard form  (after adding zero) $\tilde{C}_3 \wedge X_4 \wedge \tilde{\eta}_3$.  $D4$ is more interesting and its couplings are $( \tilde{C}_1 \wedge X_4 + \tilde{C}_{2, i} X_3\,\,^{i} ) \wedge \tilde{\eta}_5$. In the last term $c_2 \wedge X_3$ has been written as wedge product of two forms with values in U(1) contracted along the circle index.

Since the $D4$ couplings here involve a single contraction, $X_3\,\,^{i} $ is supposed to contain only odd powers of $H$.  The change of parity in $B$ is expected to come from the flip of the sign between $ \imath_v X^{(+)} $ and $ \imath_v X^{(-)} $ when written in terms of dual fields. In order to illustrate this and for sake of concreteness we shall make further simplifications. Let $D5$  worlvolume $ W_6$ be a circle bundle over $M_5$ and let the  normal bundle be trivial and $B=0$. In this situation, $T$-duality will yield  only dual $D4$ with worldvolume $M_5 \times S^1 \times \mathbb{R}^4$ and $B = a \wedge  \dd t$.  The $D4$ couplings are given by 
\begin{equation}
\label{D4}
( \tilde{C}_1 \wedge X_4 + {c}_{2} \wedge  X_3 ) \wedge \tilde{\eta}_5 \, .
\end{equation}

Recall that  the original $C_2$ gives rise a pair of  Bianchi identities $\dd F_2 = 0$ and $\dd F_3 + F_2 \wedge T = 0$ (locally $F_2 = \dd \tilde{C}_1$ and $F_3 = \dd c_2 - c_1 \wedge T$). Note that after wedging  the second with $\dd t$ we arrive at the dual BI (with $\tilde{H} =  T \wedge \dd t$): $\dd \tilde{F}_2 = 0$ and $\dd \tilde{F}_4 + \tilde{F}_2 \wedge \tilde{H} = 0$.  After integration by part arrive at \eqref{D4} yields
\begin{equation}
\label{D4*}
( \tilde{C}_1 \wedge \hat{X}_4 - {F}_{3} \wedge X_2 ) \wedge \tilde{\eta}_5 \, .
\end{equation}
As discussed,  the last term can be thought of as  a wedge product between two forms with an extra contraction  along $t$ direction. Recall that while $ X_4 $ is invariant under $T$-duality, $X_2$ is not, and $X_2(T, \imath_v H) =   \tilde{X}_2 (\imath_{v} \tilde{H}, \tilde{T})$.

Returning to a generic situation with original $H \neq 0$, we should recall that $X_4$ is averaged, and is even in $H$. So is $X_2$. The passage from $X_2$ to $\tilde{X}_2$ flips the parity in $H$. Hence in the dual coupling
\begin{equation}
\label{D4**}
( \tilde{C}_1 \wedge X_4 - \tilde{F}_{3, t} \wedge \tilde{X}_2^t) \wedge \tilde{\eta}_5 \, .
\end{equation}
(we have used a funny notation $\tilde{F}_{3, t}$ to indicate that 3 of the indices of four-form $\tilde{F}_4$ have been antisymmetrised and the fourth $t$ is contracted) the last term has odd powers of  $\tilde H$ even if the original coupling was even in $H$. This structure agrees with the string computations (note that these see only the ``linear" part of $\tilde F$). The coupling \eqref{D4**} is just the dual of familiar $I_{\tiny \mbox{D5}}$.

Hence we recover one feature of the couplings discussed in \cite{BeckRob, Ga, GaM} - on $(p+1)$-dimensional worldvolume, in addition to the standard wedged products between between a four-form made of curvatures (and $H$-flux) and $(p-3)$-form pull-back of a RR potential, contraction terms may appear as well:  instead of a four-form in curvature and $H$, we may have a vector-valued three-form contracted on the vectorial index onto a $(p-1)$ form, and a $n$-tensor valued $4-n$ form contracted into $(p-3+2n)$ RR potential. Moreover that number of contractions $n$ is correlated with the parity in $H$. Here we have seen so far only the case of $n=1$, but  in backgrounds with more than one isometries, we can see higher $n$ couplings as well.

These worldvolume couplings can be extended to the situations with more general $B$-field and backgrounds with non-trivial normal bundle. The derivation  of the couplings from the bulk action \eqref{Dbra-co} becomes more involved, but \emph{in the presence of isometries} is essentially the same as the one outlined here.

\section{Discussion}
The coupling between $D$-branes and Ramond-Ramond fields is subtle, both locally and globally. In some sense, these subtleties all have the same source: $D$-brane charges are quantised by twisted $K$-theory. As a consequence, Ramond-Ramond fields are naturally cocycles for twisted differential $K$-theory. This needs to be taken seriously: not only does this lead to an understanding of certain global problems (for example, \cite{FrMoSe} uses these considerations are used to examine subtleties in measuring Ramond-Ramond fluxes) but, as we show, doing so also explains (and predicts) local coupling terms involving derivatives of the $B$-field and the Ramond-Ramond field that do not appear in a naive approach to the coupling. The precise local expression is also important: the integrand in the coupling is not closed, so that changing the representative of the A-hat form changes the integral. 

Writing the D-brane coupling as bulk ones has a number of advantages (first of all conceptual, and then technical), including, as we have argued here, making explicit the invariance under $T$-duality. Indeed our main formula \eqref{eq:dbranebulk} is written in the bulk. However for some physical applications the knowledge of the couplings on the worldvolume may be important. Due to the failure of bulk quantities to split into tangent and normal parts when restricted to the brane, the map from the bulk to the worldvolume couplings presented here is somewhat implicit. From other side we would like to point out that in a number of applications that involve brains wrapping cycles in compact internal spaces while filling out the non-compact spacetime directions, the direct  use of \eqref{eq:dbranebulk} may prove not only possible but advantageous. These applications typically require presence of $O$-planes, whose worldvolume couplings involve the square root of the ratios of the  Hirzebruch polynomials for the tangent and normal bundles respectively up to factors of 4 \cite{SSM}. The proper understanding of the latter is tricky: \cite{DiFrMo} investigate these, along with the quantisation conditions on RR fields and the $B$-field. Their work leads naturally to a very interesting formula (Eq.~8 in \cite{DiFrMo}) which will surely underly the discussion of Ramond-Ramond couplings in the orientifold setting. We imagine that the arguments in this paper would also extend to this context, giving rise to a twisted form of the formula in the presence of $B$-fields. 

When $H$ is taken to zero, the ``contracted" terms of \cite{BeckRob, Ga, GaM} do not disappear. Indeed it is not hard to see that  even when $H$ is set to zero $X_{(s,n=2)}  (\nabla^{W/T, B}, \nu) \neq 0$ in \eqref{eq:contra}. In very local these are couplings of the form:
\begin{equation}\label{eq:verylocal}
\sim \int_{Dp} dx^{a_1}\wedge \cdots\wedge  dx^{a_{p+1}} C^{(p+1)}_{a_1\cdots a_{p-1} ij } R_{a_p i}\,^{ b k} R_{a_{p+1} j b k}
\end{equation}
where $x^{a}$ are worldvolume coordinates, $i,j,k$ denote the transverse directions and $R$ stands for the components of the Riemann tensor. These couplings are of course consistent with $T$-duality and  for two isometries can be produced by the same manipulations as in sec \ref{TDc}.
The string-theoretic derivation of these couplings  does not rely on the existence of isometries, and one might ask if they exist in general  backgrounds. From other side, they seem to contradict the known formulae for $D$-brane couplings for $B=0$!   Indeed, when $i,j,k$ are not isometry direction, the coupling \eqref{eq:verylocal} cannot be ``rotated" to the standard coupling \eqref{eq:standard} and rewritten as the $H=0$ limit of the bulk coupling \eqref{eq:dbranebulk}. Our guess is that if the background has no isometries, all couplings on the worldvolume not accounted the restriction of \eqref{eq:dbranebulk} using the modified Riemann-Roch theorem  \ref{th:twistrr}  are due to  the ambiguity of the slitting principle and hence should amount to exact shifts of the square root of $\hat A$. We have no proof of this right now, but think it is an interesting question.

In a different direction, it is natural to seek contact with the picture of $D$-branes as generalised submanifolds in the sense of~\cite{Gualt}. Just as a $B$-field induces a generalised geometry, the trivialisation of the $B$-field on the support of the $D$-brane is what is needed give the support the structure of a generalised submanifold.   Taking this structure into account may have some benefits: notably, for our purposes, the geometry may be seen as taking place on a deformation of the sum the tangent and co-normal bundles of the support of the $D$-brane. It may be expected that a proper account of geometric index theory in the generalised geometric setting would define push-forwards from generalised sub-manifolds, and that the correct ``A-hat" form in the ``generalised geometric" Riemann-Roch for such pushforwards would thus be a deformation of the ratio of the A-hat forms of the tangent and normal bundles. In turn, this deformation would surely form the basis of a refinement of the arguments in this paper, and, in particular would hopefully allow the formulation of an explicit expression for the $D$-brane/Ramond-Ramond field coupling on the worldvolume.
\bibliographystyle{amsplain}
\bibliography{dbrane}
\end{document}